\documentclass[prb,aps,groupedaddress,superscriptaddress, twocolumn,toc=flat]{revtex4-1}
\usepackage[colorlinks=true,linkcolor=blue,urlcolor=blue,citecolor=blue]{hyperref}
\usepackage[utf8x]{inputenc}
\usepackage{color}
\usepackage{bbm}
\usepackage{amsfonts,amsmath,amssymb,stmaryrd}
\usepackage{graphicx}
\usepackage{subfigure}  
\usepackage{bbm}
\usepackage{natbib}
\usepackage{epsfig}
\usepackage{mathrsfs}
\usepackage{verbatim}
\usepackage{centernot}
\usepackage{ulem}
\usepackage{nicefrac}
\usepackage{array}
\usepackage{subfigure}

\renewcommand{\figurename}{\textbf{Fig.}}
\makeatletter
\renewcommand{\fnum@figure}{\textbf{Fig.} \textbf{\thefigure}}
\makeatother

\begin{document}

 \normalem

 \title{Geometric Frustration Assisted Kinetic Ferromagnetism in Doped Mott Insulators}

\author{Qianqian Chen}
\affiliation{Kavli Institute for Theoretical Sciences, University of Chinese Academy of Sciences, Beijing 100190, China}
\author{Shuai A. Chen}
\affiliation{Max Planck Institute for the Physics of Complex Systems, N\"{o}thnitzer Stra{\ss}e 38, Dresden 01187, Germany}
\author{Zheng Zhu}
\email{zhuzheng@ucas.ac.cn}
\affiliation{Kavli Institute for Theoretical Sciences, University of Chinese Academy of Sciences, Beijing 100190, China}

\begin{abstract}
Understanding ferromagnetism mechanism in doped Mott insulators on frustrated lattices remains challenging at intermediate coupling and finite doping. Here, we study the itinerant ferromagnetism and propose its mechanism in doped Mott insulators on a geometrically frustrated triangular lattice. Using large-scale density matrix renormalization group (DMRG) and unrestricted Hartree-Fock mean-field methods, we reveal that itinerant ferromagnetism appears at intermediate coupling ($10\lesssim U\ll\infty$) near 50\% electron doping in the triangular-lattice Hubbard model. By analyzing all microscopic hopping processes, we find that doublon-singlon  exchange alone drives the fully polarized ferromagnetism and uncovers the particle-hole asymmetry. We also establish the magnetic phase diagram and compare local spin correlations with recent experiments. Random phase approximation and DMRG calculations consistently confirm that the ferromagnetism persists when $SU(2)$ symmetry is explicitly broken by magnetic anisotropy. These results clarify a microscopic route to itinerant ferromagnetism at intermediate coupling and finite doping in doped Mott insulators.

\end{abstract}

\maketitle

~\\
\textsf{\textbf{\large Introduction.}}
~\\

It has been a long-standing issue in physics to understand the ferromagnetism mechanism in strongly correlated systems. As the fundamental theoretical model for diverse physics of these systems, the Hubbard model is initially formulated to understand the origin of ferromagnetism \cite{Gutzwiller1963,Hubbard1963}.
Given the lack of explicit or effective magnetic interactions favoring ferromagnetic order, ferromagnetism can only arise from the delicate interplay between kinetic energy and on-site repulsion $U$. To date, rigorous theoretical results on the ferromagnetism mechanism in the Hubbard model have only been achieved in specific limits~\cite{Tasaki1998, Tasaki1998Review, Vollhardt1999}. 
The Stoner criterion~\cite{Stoner1938} states that ferromagnetism occurs when the product of the density of states at the Fermi energy $D_{\mathrm{F}}$ and on-site Coulomb repulsion $U$ satisfies the condition $U D_{\mathrm{F}}>1$. However, this criterion is derived from the Hartree-Fock approximation and is known to often overpredict ferromagnetic tendencies \cite{Vollhardt1996, Vollhardt1999}.
Two rigorous outcomes are Nagaoka ferromagnetism~\cite{Thouless1965,Nagaoka1966}, arising from kinetic energy minimization of a single doped charge at $U=\infty$, and flat-band ferromagnetism~\cite{Mielke1991b,Tasaki1992} with dispersionless lowest energy bands. 
Over the past several decades, it has attracted substantial efforts to explore ferromagnetism beyond those well-known limits~\cite{Riera1989, Basile1990,Emery1990, Shastry1990,Mielke1991,Mielke1992, Hanisch1995, Wurth1996, Brunner1998, Ulmke1998, Becca2001, Merino2006, ChiaChenChang2010, SoonYongChang2011, Carleo2011, LiLiu2012, GangLi2014, Schlomer2023,Davydova2023, Samajdar2023, YuchiHe2023, Morera2024, Lorenzo2024}, particularly for the square lattices~\cite{Riera1989, Basile1990, Emery1990, Shastry1990, Wurth1996, Brunner1998, Carleo2011, LiLiu2012}. {However, understanding the microscopic mechanism for ferromagnetism in geometrically frustrated triangular lattices at both intermediate Coulomb repulsion {($10\lesssim U\ll\infty$)} and finite doping $\delta$ remains an open issue with fundamental challenge~\cite{Merino2006,Samajdar2023,YuchiHe2023, Morera2024,Samajdar2024PRB}.}

More recently, the transition metal dichalcogenides (TMD) moiré materials~\cite{YanhaoTang2020, Ciorciaro2023, Anderson2023, Seifert2024} and the quantum simulators in optical lattices~\cite{YangJin2021,MuqingXu2023, Lebrat2024,Prichard2023,ChinCheng2010,Lewenstein2012}  have significantly advanced this issue. The essential physics of these experimental platforms could be captured by the triangular-lattice Hubbard model. With unprecedented control over doping and coupling strength, ferromagnetism has been observed  at an intermediate coupling and large electron doping. These experimental advancements also necessitate theoretical study on the mechanisms and stability of the observed ferromagnetism in triangular lattices.

Unlike the intensively studied hole-doped case, experimentally observed ferromagnetism occurs at the electron-doped side, the physics of which could be quite different due to the particle-hole asymmetry~\cite{Zhu2023SC,ShuaiChen2022SCBA,MuqingXu2023, Prichard2023,GuanhuaHuang2023}.
In the limit of nearly half filling $n$$\to$$1$ ($n$ denotes electron density per site) and full filling $n$$\to$$2$, the Nagaoka-type~\cite{Hanisch1995,ShuaiChen2022SCBA, Morera2023, Davydova2023, Schlomer2023, KyungminLee2023,Morera2024} and Müller-Hartmann-type ferromagnetism~\cite{ChengshuLi2023} have been proposed, respectively. However, the ferromagnetism mechanism for intermediate density around $n=3/2$,  {which is significantly away from half-filling and full-filling}, remains an open issue. At $n=3/2$, 
weak interactions would trigger instabilities towards magnetic orders with nesting wave vectors~\cite{Martin2008,Akagi2010}. However, when the weak-coupling theory fails at the strong-coupling regime~\cite{Pasrija2016}, it is still a much-needed endeavor to identify the magnetism and its mechanism,  {particularly for the intermediate coupling regions}.

\begin{figure*}[tpb]
\begin{center}
\includegraphics[width=0.8\textwidth]{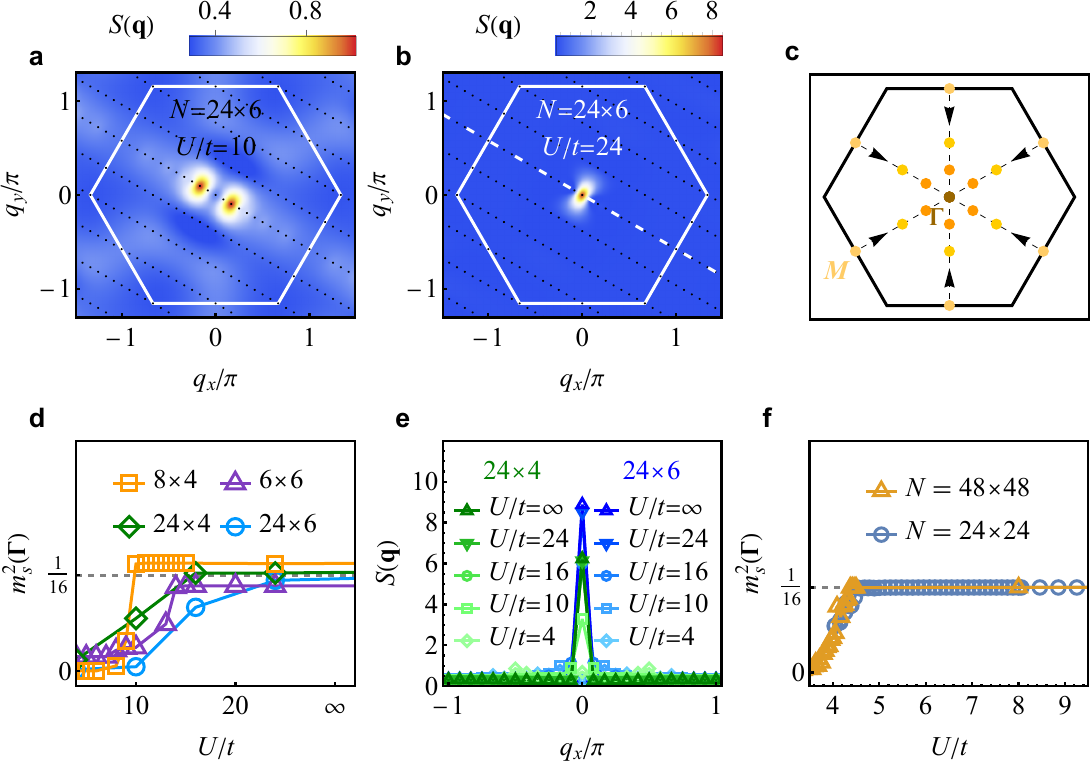}
\end{center}
\par
\caption{
\textbf{$\vert$
The static spin structure factor $\boldsymbol{ S(\mathbf{q})}$ as a function of {coupling strength} $\boldsymbol{U/t}$ at electron doping $\boldsymbol{\delta=1/2}$.
}
(\textbf{a})-(\textbf{b}) The contour plot of $S(\mathbf{q})$ for cylinders  at  $U/t=10$ (\textbf{a}), and $U/t=24$ (\textbf{b}).  The black dots represent the accessible momenta in the Brillouin zone (denoted as a white hexagon), and interpolation has been applied in the contour plot. We remark that the cylindrical geometry adopted in {the density matrix renormalization group (DMRG) simulation} breaks the lattice's rotational symmetry. 
(\textbf{c}) Schematic of the evolution of the  $\mathbf{q}_0$  with increasing $U/t$, indicated by the color gradient from light to dark along the black dashed arrows. Here, $\mathbf{q}_0$ denotes the momenta where $S(\mathbf{q})$ reaches its maximum, within the Brillouin zone represented by a hexagon. Rotational symmetry of triangular lattices is assumed in this schematic for clarity.
(\textbf{d}) Ferromagnetic squared order parameter $m_s^2(\mathbf{\Gamma})$ as a function of $U/t$. For larger $U/t$, the peak of $S(\mathbf{q})$ stabilizes at momentum $\mathbf{\Gamma}$ with {a size-independent saturation value}, indicating the fully polarized ferromagnetism. 
(\textbf{e}) Line-cut plot of $S(\mathbf{q})$ along the momentum path passing through the $\mathbf{\Gamma}$ point in (\textbf{c}), depicted as a white dashed line in (\textbf{b}). 
(\textbf{f}) Squared order parameter $m_s^2(\mathbf{\Gamma})$ from unrestricted Hartree-Fock calculations with periodic boundary conditions.  }
\label{Fig_FMPropertySq}
\end{figure*}

\begin{figure}[tbp]
\begin{center}
\includegraphics[width=0.48\textwidth] {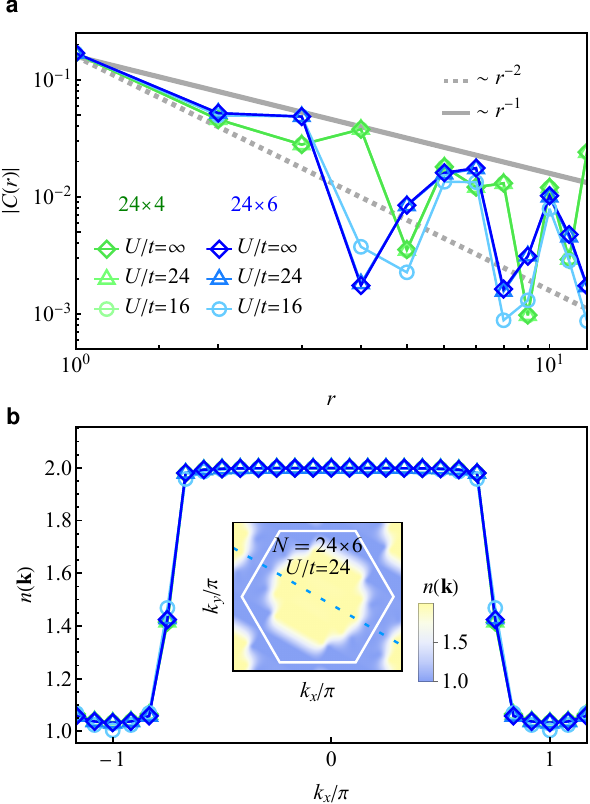}
\end{center}
\par
\caption{
\textbf{$\vert$ The single-particle propagator $\boldsymbol{ C(r)}$  {at separation $\boldsymbol{r}$} and electron momentum distribution $\boldsymbol{ n(\mathbf{k})}$ at  electron doping $\boldsymbol{\delta=1/2}$.}
(\textbf{a}) Single-particle propagator $|C(r)|$ for different $U/t$, {where $U$ is the on-site Hubbard interaction, and $t$ is the nearest-neighbor hopping amplitude}. For comparison, the dashed and solid gray lines correspond to $|C(r)|\sim r^{-2}$ and $|C(r)|\sim r^{-1}$, respectively.
(\textbf{b}) Line-cut plot of electron momentum distribution $n(\mathbf{k})$ along a specific momentum path that traverses the momentum $\mathbf{\Gamma}$, depicted as a blue dashed line in its inset. The inset shows the contour plot of $n(\mathbf{k})$, {the white hexagon denotes the first Brillouin zone, and the colour bar encodes $n(\mathbf{k})$}. 
}
\label{Fig_FMPropertyCharge}
\end{figure}

Motivated by the above, we explore the ferromagnetism and its underlying mechanism in the triangular-lattice Hubbard model from intermediate to infinite interactions near electron doping $\delta=1/2$ (i.e., $n=3/2$, where the doping concentration is defined by $\delta\equiv n-1$). Using large-scale density matrix renormalization group (DMRG) and unrestricted Hartree-Fock (UHF) methods, we first present the emergence of  ferromagnetic ground state with metallic charge behavior at $n=3/2$ as  the on-site Coulomb repulsion $U$  increases from intermediate to strong and ultimately infinite values. Among distinct microscopic hopping processes inherent {in a decomposition of the intermediate-$U$ Hubbard model}, we reveal that {only the doublon-singlon exchange---assisted by the geometric frustration of the lattice---contributes to the fully polarized ferromagnetism.} {We further offer stronger evidence for this conclusion}
by selectively deactivating each hopping process and verifying their individual impacts on itinerant ferromagnetism. Additionally, we demonstrate that itinerant ferromagnetism around $\delta=1/2$ persists within a specific range of electron doping and find the local spin correlations at zero temperature resemble those observed in a recent experiment~\cite{MuqingXu2023}.
{Finally, we confirm that ferromagnetism is robust against the explicit breaking of $SU(2)$ symmetry in the Hubbard model through the introduction of anisotropy.}

~\\
\textsf{\textbf{\large RESULTS}}
~\\
\textbf{Itinerant Ferromagnetism at Intermediate $U$ and Large Electron Doping.}
In this work, we study the ground-state properties of the triangular-lattice Hubbard model at intermediate coupling strength. The many-body Hamiltonian reads
\begin{equation}\label{eq:model}
H=-t \sum_{\langle \mathbf{i} \mathbf{j}\rangle, \sigma}\left(c_{\mathbf{i} , \sigma}^{\dagger} c_{\mathbf{j} ,\sigma}+\text {H.c.}\right)+U \sum_\mathbf{i} n_{\mathbf{i} , \uparrow} n_{\mathbf{i}, \downarrow}.
\end{equation}
Here, $t>0$ and $U>0$ are the nearest-neighbor (NN) hopping amplitude and the on-site Coulomb repulsion energies. The operators $c_{\mathbf{i}, \sigma}^{(\dagger)}$ ($c_{\mathbf{i}, \sigma}$) creates (annihilates) an electron on site $\mathbf{i}$ with spin $\sigma=\uparrow,\downarrow$, and $n_{\mathbf{i} ,\sigma}\equiv c_{\mathbf{i} ,\sigma}^{\dagger} c_{\mathbf{i}, \sigma}$ is the number operator. The summation of bonds $\langle\cdots\rangle$ runs over all NN bonds in the triangular lattice.  The doping concentration is represented by $\delta = n-1$,  with $n$ indicating the electron density per site. Here, $\delta=0$ corresponds to the half-filling, and $\delta>0$ ($\delta<0$) is the electron-doped (hole-doped) case. Here, we fix $t=1$.

We first identify the itinerant ferromagnetic phase at electron doping $\delta=1/2$ and intermediate coupling strength in the triangular-lattice Hubbard model using DMRG and UHF methods. {Unless stated otherwise, the figures in the following show DMRG data obtained on $L_x \times L_y$ cylindrical lattices with primitive vectors $\boldsymbol{e}_x = (1, 0)$ and $\boldsymbol{e}_y = (1/2, \sqrt{3}/2)$, open boundary conditions along $L_x$ (length), and periodic along $L_y$ (circumference). UHF is applied with periodic boundary conditions in both directions.
We remark that our largest DMRG simulations are performed on systems with width up to ${L_y}=6$ and ${L_x}\leq24$. These system sizes (i) significantly suppress finite-size effects and (ii) are crucial for clearly distinguishing ferromagnetism from the incommensurate spin density wave (iSDW) correlations.}

To identify the nature of the ground state in the spin and charge channels at electron doping $\delta=1/2$, we examine the evolution of the magnetic and electronic structure across a range of interaction strengths, from finite to infinite $U/t$, as depicted in Fig.~\ref{Fig_FMPropertySq} and Fig.~\ref{Fig_FMPropertyCharge}.
The magnetic order is signaled by peaks at certain wave vectors $\mathbf{q}=\mathbf{q}_0$ in the static spin structure factor $S(\mathbf{q})$, defined by $S(\mathbf{q})\equiv\sum_{\mathbf{i},\mathbf{j}} 
\langle {\mathbf{S}_\mathbf{i}\cdot \mathbf {S}_\mathbf{j}} \rangle 
e^{i \mathbf{q}\cdot (\mathbf{i}-\mathbf{j})}/N$.

We first focus on the evolution of the momenta $\mathbf{q}_0$
as a function of interaction strength $U/t$.
In the noninteracting limit ($U/t=0$), electron doping $\delta=1/2$ corresponds to the Van-Hove filling at which a perfect nesting of the Fermi surface exists.
At weak coupling side $U/t\to0$, it is expected that nesting leads to peaks of $S(\mathbf{q})$ locating at the $\mathbf{M}$ points in the Brillouin zone {[as labeled in Fig.~\ref{Fig_FMPropertySq}c]}. Previous weak-coupling theory \cite{Martin2008} proposes that the candidate ground state exhibits noncoplanar chiral magnetic order.
When the weak coupling assumption is no longer applicable at intermediate to strong interaction strengths $U/t$, using DMRG we numerically observe that other magnetic ordering vectors distinct from the nesting vectors emerge, as shown in Figs.~\ref{Fig_FMPropertySq}a.   Interestingly, as $U/t$ increases, the locations of these peaks in $S(\mathbf{q})$ shift towards the $\mathbf{\Gamma}$ point {[as labeled in Fig.~\ref{Fig_FMPropertySq}c]}, which is also consistent with the recent Hartree-Fock analysis \cite{YuchiHe2023}.  Upon surpassing a critical strength $U_c/t$ toward the limit $U/t=\infty$, the peak of $S(\mathbf{q})$ becomes fixed at $\mathbf{\Gamma}$ point, as illustrated in  Fig.~\ref{Fig_FMPropertySq}b, d, indicating the establishment of a stabilized ferromagnetic phase in the strong coupling regime.
In Fig.~\ref{Fig_FMPropertySq}c, we schematically plot such evolution of the peaks in $S(\mathbf{q})$ with increasing $U/t$, as visualized by the dashed arrow from momenta $\mathbf{M}$ to $\mathbf{\Gamma}$, and the additional supporting numerical data are provided in the Supplementary Note 1.
Notably, after the peak of $S(\mathbf{q})$ shifts to the $\mathbf{\Gamma}$ point, its height gets larger in a finite range of increasing $U/t$ [see Fig.~\ref{Fig_FMPropertySq}e], until reaching its maximum value. This signals the evolution from a partially-polarized to a fully-polarized phase.
We therefore keep track of the squared order parameter $m_s^2(\mathbf{\Gamma}) \equiv S(\mathbf{\Gamma})/N$ as a function of $U/t$. As illustrated by Fig.~\ref{Fig_FMPropertySq}d, when increasing $U/t$, $m_s^2(\mathbf{\Gamma})$ gradually increases until plateauing, signaling the establishment of a stabilized ferromagnetic phase at intermediate $U/t$. The saturated value matches the ferromagnetic squared order parameter $m_s^2(\mathbf{\Gamma})\sim\delta_{\mathbf{q},\mathbf{\Gamma}}/16$ for a fully polarized state at electron doping $\delta=1/2$.

Having established the ferromagnetism in the spin channel, we further examine the electronic properties in the charge channel. 
We compute the single-particle propagator $
 C(r)\equiv\sum_{\sigma}\langle c_{\mathbf{i}_0,\sigma}^\dagger c_{{\mathbf{i}_0+r\boldsymbol{e}_x},\sigma}\rangle 
$, and the momentum distribution of the electrons $
  n(\mathbf{k})\equiv\sum_{\mathbf{i}, \mathbf{j} , \sigma}\langle c_{\mathbf{i}, \sigma}^{\dagger} c_{\mathbf{j}, \sigma}^{}\rangle e^{i \mathbf{k} \cdot(\mathbf{i}-\mathbf{j})}/N
$. 
Fig.~\ref{Fig_FMPropertyCharge}a illustrates the power-law decay of $|C(r)|$ with an exponent approximately equal to 1, identifying the gapless nature of the electrons in the ferromagnetic phase. In the inset of Fig.~\ref{Fig_FMPropertyCharge}b, the contour plot of $n(\mathbf{k})$ exhibits an abrupt change from 2 to 1 at certain momenta, as more clearly illustrated by the line-cut plot of $n(\mathbf{k})$ through the momentum $\mathbf{\Gamma}$ in the main panel of Fig.~\ref{Fig_FMPropertyCharge}b. This abrupt change indicates the presence of a well-defined Fermi surface, with the position of this sudden change characterizing the Fermi momenta. These observations are robust for the strength of intermediate-to-infinite interactions $U/t$ and different system sizes $N$, suggesting that the electrons in the ferromagnetic phase for both intermediate $U/t$ and infinite $U/t$ limit are itinerant.

To confirm the robustness of ferromagnetism in the thermodynamic limit, we further employ the unrestricted Hartree-Fock (UHF) mean-field calculations. 
In the UHF method, under an uncorrelated state ansatz, both the on-site densities $\langle n_{\mathbf i,\sigma}\rangle$ and the spin-flips $\langle S_{\mathbf i}^{-}\rangle$ and $\langle S_{\mathbf i}^{+}\rangle$ are involved to factorize the Hubbard interaction.
Here $S_{\mathbf{i}}^{+}=c_{\mathbf{i},\uparrow}^{\dagger} c_{\mathbf{i}\downarrow}^{}$ and $S_{\mathbf i}^{-}=c_{\mathbf{i},\downarrow}^{\dagger} c_{\mathbf{i},\uparrow}^{}$ are spin ladder operators.  Then the Hartree-Fock factorizes the Hamiltonian to
\begin{align}
H_\mathrm{HF}= & -\sum_{\langle \mathbf{ij}\rangle}(t+\mu\delta_{\mathbf{ij}})c_{\mathbf i,\sigma}^{\dagger}c_{\mathbf j,\sigma}^{}\notag \\
&+\sum_{\mathbf i}U[\langle n_{\mathbf i,\uparrow}\rangle n_{\mathbf i,\downarrow}+\langle n_{\mathbf i,\downarrow}\rangle n_{\mathbf i,\uparrow}-\langle S_{\mathbf i}^{+}\rangle S_{\mathbf i}^{-}-\langle S_{\mathbf i}^{-}\rangle S_{\mathbf i}^{+}]\nonumber \\
 & +\sum_{\mathbf i}U\left[-\langle n_{\mathbf i,\uparrow}\rangle\langle n_{\mathbf i,\downarrow}\rangle+\langle S_{\mathbf i}^{-}\rangle\langle S_{\mathbf i}^{+}\rangle\right],
\end{align}
where the chemical potential $\mu$ controls the electron
doping $\delta=1/2$. The mean-field parameters are expectation values under the ground state $\vert\Omega\rangle$ of the $H_\mathrm{HF}$,
\begin{equation}
\langle n_{\mathbf i,\sigma}\rangle\equiv\langle\Omega\vert n_{\mathbf i,\sigma}\vert\Omega\rangle,\quad\langle S_{\mathbf i}^{\pm}\rangle\equiv\langle\Omega\vert S_{\mathbf i}^{\pm}\vert\Omega\rangle,
\end{equation}
and the on-site densities are subject to the particle number constraint $
\sum_{\mathbf i}[\langle n_{\mathbf i,\uparrow}\rangle+\langle n_{\mathbf i,\downarrow}\rangle]=N(1+\delta)$ with $N$ being the number of sites.  There are a total $4N$ variational parameters, which can be iteratively solved by the
self-consistent equations
\begin{align}
\langle n_{\mathbf i,\sigma}\rangle & =\sum_{n}f_{FD}(E_{n})\langle\psi_{n}\vert n_{\mathbf i,\sigma}\vert\psi_{n}\rangle\\
\langle S_{\mathbf i}^{\pm}\rangle & =\sum_{n}f_{FD}(E_{n})\langle\psi_{n}\vert S_{\mathbf i}^{\pm}\vert\psi_{n}\rangle
\end{align}
where $\vert\psi_{n}\rangle$ is an eigenstate of $H_\mathrm{HF}$ with $H_\mathrm{HF}\vert\psi_{n}\rangle=E_{n}\vert\psi_{n}\rangle$
and $f_{FD}(E_{n})$ is the Fermi-Dirac distribution. 
For a stable convergence, we utilize the direct inversion in
the iterative subspace (DIIS) method~\cite{Pulay1980}, and the convergence criterion for the energy is set to $10^{-8}t$.

As shown in Fig.~\ref{Fig_FMPropertySq}f, the squared order parameter $m_s^2(\mathbf{\Gamma})$ increases gradually as $U/t$ becomes larger, and finally a full polarization state occurs at $U\simeq 4.5t$ independently of system size. 
This aligns with Fig.~\ref{Fig_FMPropertySq}d and is also consistent with the recent Hartree-Fock analysis \cite{YuchiHe2023}, indicating robust fully-polarized ferromagnetism in the strong coupling regime.

\begin{figure}[tp]
\begin{center}
\includegraphics[width=0.51\textwidth]{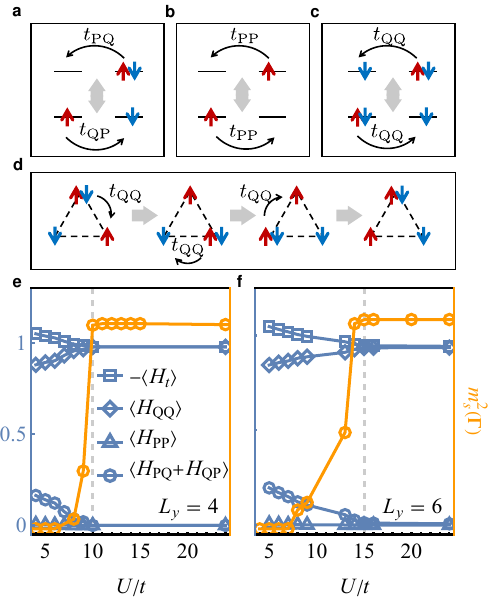}
\end{center}
\par
\caption{
\textbf{$\vert$ Microscopic hopping processes in the Hubbard model and their kinetic energy contributions for intermediate {coupling} $\boldsymbol{ U/t}$.}
(\textbf{a})-(\textbf{c}) Schematic illustration of nearest-neighbor hopping processes described in Eq.~\eqref{eq:SepHoppingHt}.
(\textbf{d}) Schematic illustration of ring-exchange coupling.
(\textbf{e}),(\textbf{f}) The kinetic energies (in blue) of different microscopic hopping processes as functions of $U/t$ for {lattice width} $L_y=4$ in (\textbf{e}) and $L_y=6$ in (\textbf{f}). Ferromagnetic squared order parameter $m_s^2(\mathbf{\Gamma})$ (in orange) is plotted for comparison. 
}
\label{Fig_Cnnsum}
\end{figure}

~\\
\textbf{Microscopic mechanism: geometrical frustration assisted kinetic ferromagnetism.}
In the above, we have identified the emergence of itinerant ferromagnetism with increasing $U/t$ at $\delta=1/2$. 
To understand the underlying mechanism for the itinerant ferromagnetism, we examine the distinct microscopic hopping processes encompassed in the hopping term $H_t$ of the intermediate-$U$ Hubbard model, 
as schematically illustrated in Figs.~\ref{Fig_Cnnsum}a, b, c. 
We introduce the projection operator $\hat{Q}_{\mathbf{i},{\sigma}}\equiv n_{\mathbf{i}, {\sigma}}$ and  $\hat{P}_{\mathbf{i},{\sigma}}\equiv1-\hat{Q}_{\mathbf{i},{\sigma}}$ to decompose the hopping term $H_t$:
\begin{equation}\label{eq:SepHoppingHt}
    {\small
        \begin{split}
    H_t=&-\sum_{\langle \mathbf{i} \mathbf{j}\rangle, \sigma} \bigl(
    t_\mathrm{PQ} \hat{P}_{\mathbf{i},\bar{\sigma}} c_{\mathbf{i},\sigma}^{\dagger} c_{\mathbf{j},\sigma} \hat{Q}_{\mathbf{j},\bar{\sigma}}
    + t_\mathrm{QP} \hat{Q}_{\mathbf{i},\bar{\sigma}} c_{\mathbf{i},\sigma}^{\dagger} c_{\mathbf{j},\sigma} \hat{P}_{\mathbf{j},\bar{\sigma}}\\
    &+ t_\mathrm{PP} \hat{P}_{\mathbf{i},\bar{\sigma}} c_{\mathbf{i},\sigma}^{\dagger} c_{\mathbf{j},\sigma} \hat{P}_{\mathbf{j},\bar{\sigma}}
    + t_\mathrm{QQ} \hat{Q}_{\mathbf{i},\bar{\sigma}} c_{\mathbf{i},\sigma}^{\dagger} c_{\mathbf{j},\sigma} \hat{Q}_{\mathbf{j},\bar{\sigma}}+ \text{H.c.} \bigr)\\
    \equiv&-\sum_{\langle \mathbf{i} \mathbf{j}\rangle, \sigma} \bigl[
    t_\mathrm{PQ} (H_{\mathrm{PQ}} + H_{\mathrm{QP}})+t_\mathrm{PP} H_{\mathrm{PP}}+t_\mathrm{QQ} H_{\mathrm{QQ}}
    \bigr].
        \end{split}
    }
\end{equation}
Consequently,   $H_t$ includes NN hopping between 
(i) a doubly occupied site and an empty site with hopping amplitude $t_\mathrm{PQ}$, or two singly occupied sites with hopping amplitude $t_\mathrm{QP}$ [see Fig.~\ref{Fig_Cnnsum}a]; 
(ii) a singly occupied site and an empty site with hopping amplitude $t_\mathrm{PP}$ [see Fig.~\ref{Fig_Cnnsum}b], i.e., singlon hopping; 
(iii) a doubly occupied and a singly occupied site with hopping amplitude $t_\mathrm{QQ}$ [see Fig.~\ref{Fig_Cnnsum}c], i.e., the doublon-singlon exchange. Given that case (i) describes Hermitian conjugate processes for an isotropic case,  
we consider $t_\mathrm{PQ}=t_\mathrm{QP}$.  
The hopping term in the Hubbard model~\eqref{eq:model} corresponds to the isotropic limit $t_\mathrm{PQ}=t_\mathrm{QP}=t_\mathrm{PP}=t_\mathrm{QQ}=t$. 
 {
In contrast to the large $U$ limit, where the $t$-$J$ model effectively describes the Hubbard model with only the $t_\mathrm{QQ}$ hopping term in the no-empty-occupancy Hilbert space for electron doping ($n>1$) \cite{ShuaiChen2022SCBA}, intermediate $U$ regimes are generally regarded as involving multiple kinetic processes that cannot be reduced to the $t$-$J$ model alone~\cite{Spalek2007,KeYang2024}.
However, it remains unclear whether ferromagnetism at intermediate $U$ is induced by a single kinetic process or by the cooperative effect of several processes. 
}

The decomposition \eqref{eq:SepHoppingHt} enables us to examine the individual contributions of each microscopic hopping to itinerant ferromagnetism.  As shown in Figs.~\ref{Fig_Cnnsum}e, f for $L_y=4, 6$ DMRG cylinders, before entering the fully polarized region where $m_s^2(\boldsymbol{\Gamma})$ saturates, multiple hopping processes have nonzero expectation values, suggesting that they contribute to the ground-state kinetic energy. In contrast, when increasing $U/t$  {to intermediate coupling}, only $\langle H_\mathrm{QQ}\rangle$ is nonzero in the fully polarized region.  {We remark that, unlike the infinite $U$ or strong-coupling limits, the result in Figs.~\ref{Fig_Cnnsum}e, f is particularly unexpected for intermediate coupling  ($U/t \sim 10$)  where multiple kinetic processes are typically significant.} 
These  observations illustratively suggests that,  among distinct hopping processes {in the decomposition \eqref{eq:SepHoppingHt}}, the doublon-singlon exchange 
{dominantly contribute to the kinetic energy of fully polarized ferromagnetism.}
While this is conceptually similar to Nagaoka’s ferromagnetism, 
the conditions for the emergence of the ferromagnetism---intermediate interaction ($U \gtrsim 10$) and intermediate doping ($\delta = 1/2$)---significantly violate the strict preconditions of Nagaoka's theorem. 
{In order to compare to the standard Nagaoka regime (single electron doping and $U=\infty$), we repeated the Fig.~\ref{Fig_Cnnsum}e analysis for a single electron doped $8 \times 4$ cylinder and found that, even at $U / t=260$, the ground state remains only partially polarized. Nonetheless, similar behavior persists: both $\langle H_{\mathrm{PQ}}+H_{\mathrm{QP}}\rangle$ and $\langle H_{\mathrm{QQ}}\rangle$ contribute at intermediate couplings, while the doublon-singlon exchange term $\langle H_{\mathrm{QQ}}\rangle$ steadily dominates as $U / t$ increases.
This can be naturally extrapolated to the $U\to \infty$ limit, where the doublon-singlon exchange $\langle H_{\mathrm{QQ}} \rangle$ becomes the sole active kinetic process in the fully polarized ferromagnetic state.}
We then focus on doublon-singlon exchange and show that the doublon-assisted ring-exchange [see Fig.~\ref{Fig_Cnnsum}d] leads to the ferromagnetism, unlike the antiferromagnetism typically driven by superexchange.

\begin{figure}[!t]
\begin{center}
\includegraphics[width=0.49\textwidth]{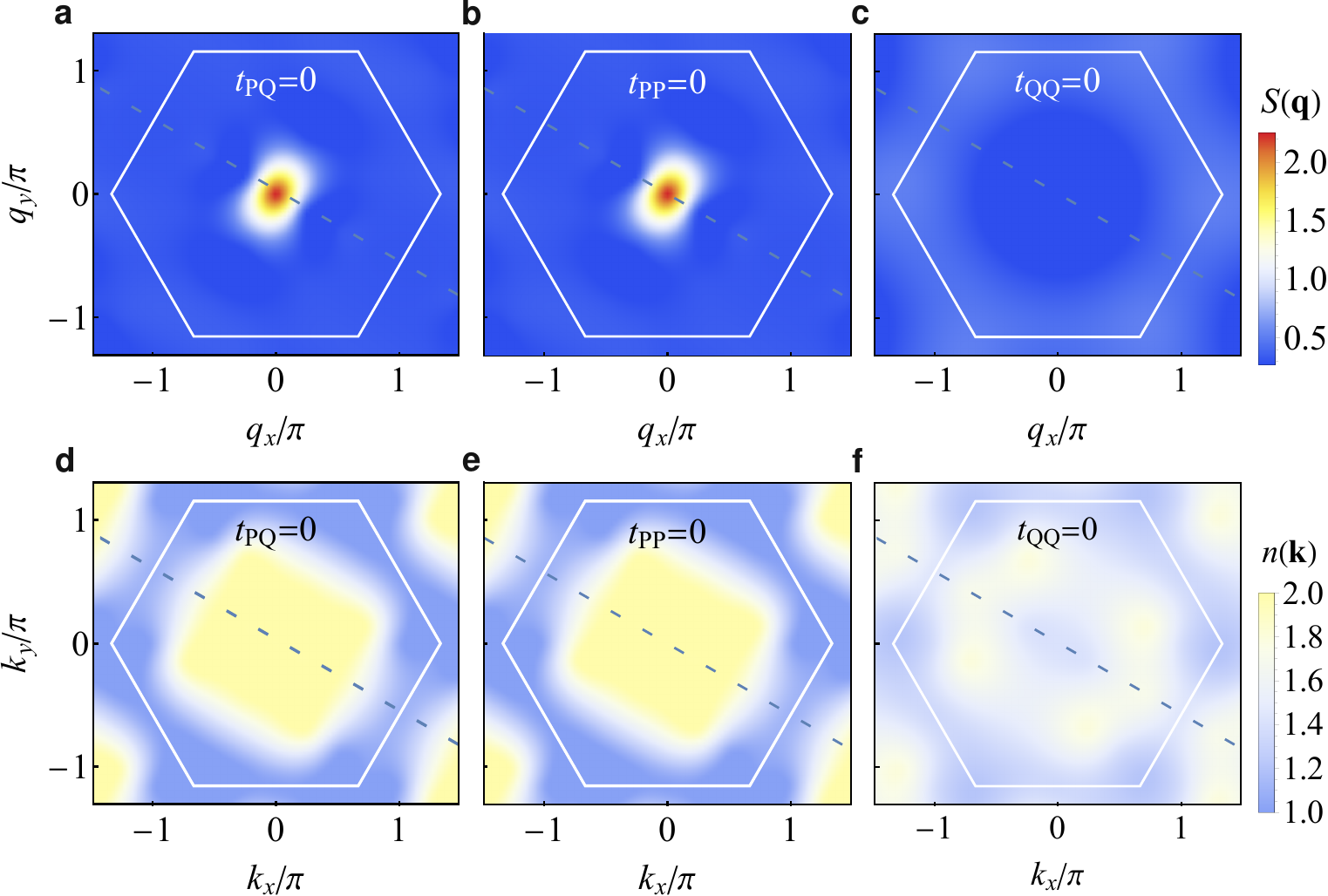}
\end{center}
\par
\caption{
\textbf{$\vert$
Impact of the microscopic hopping processes on itinerant ferromagnetism in the Hubbard model.}
(\textbf{a})-(\textbf{f}) Static spin structure factor $S(\mathbf{q})$ ((\textbf{a})-(\textbf{c})) and the momentum distribution $n(\mathbf{k})$ ((\textbf{d})-(\textbf{f})) obtained by independently turning off each hopping process: (\textbf{a}),(\textbf{d}) $t_\mathrm{PQ}=0$, (\textbf{b}),(\textbf{e}), $t_\mathrm{PP}=0$, (\textbf{c}),(\textbf{f}), $t_\mathrm{QQ}=0$. Other hopping terms in Eq.~\eqref{eq:SepHoppingHt} are set to $t$. The white hexagon indicates the first Brillouin zone; interpolation is applied to contours. 
{Here, we fix the system size as $8\times 4$ cylinders, the electron doping $\delta=1/2$, and the interaction strength $U=10$.}
}
\label{Fig_4Hopping}
\end{figure}

The effective NN spin interactions  stem from two kinds of virtual hopping processes at finite electron doping: (i) the superexchange $t^2/U$ favoring antiferromagnetic order, 
 {
(ii) the doublon-assisted ring-exchange $t^3/\Delta E^2$   favoring either antiferromagnetic or ferromagnetic order depending on the sign of hopping $t$~\cite{Thouless1965, Tasaki1998}, where $\Delta E$ refers to the energy difference between the initial and intermediate states.
Therefore, the effective spin interaction is $\sim (t^2/U+{t^3}/{\Delta E^2}) \mathbf{S}_i\cdot \mathbf{S}_j$.   This formulation reveals a fundamental difference between hole and electron doping in triangular lattices.  
For hole-doping ($t>0$), both hopping processes favor antiferromagnetism; while for electron-doping ($t<0$), the ring-exchange virtual process promotes ferromagnetism and finally dominates over the superexchange coupling in the large and infinite $U$ limit. 
}
Notably, square lattices lack such particle-hole asymmetry  due to the positive ring-exchange coupling $\sim t^4$.
This distinction highlights the important properties of geometrically frustrated lattices like the triangular lattice, where kinetic energy and geometric frustration work synergistically to foster itinerant ferromagnetism.
In addition to the above perturbation analysis, we also provide an alternative explanation grounded in the symmetry of wave functions to understand the ferromagnetism origin (see Supplementary Note 2).

{
To show stronger evidence that the doublon-singlon exchange process drives ferromagnetism}, we independently deactivate  
each hopping amplitude while keeping the others unchanged, and compare the resulting ground states with the standard Hubbard model~\eqref{eq:model}.  
In this manner, the specific hopping process whose deactivation eliminates itinerant ferromagnetism is decisive for ferromagnetism. 
Figures~\ref{Fig_4Hopping} reveal that only switching off $t_\mathrm{PQ}$ or $t_\mathrm{PP}$ does not significantly affect the ferromagnetism, as the resulting  $S(\mathbf{q})$ [see Figs.~\ref{Fig_4Hopping}a, b] and $n(\mathbf{k})$  [see Figs.~\ref{Fig_4Hopping}d, e] remain similar to those of the Hubbard model \eqref{eq:model} [see Fig.~\ref{Fig_FMPropertySq}e and Fig.~\ref{Fig_FMPropertyCharge}b]. 
Conversely, switching off $t_\mathrm{QQ}$ eliminates ferromagnetic signatures [Figs.~\ref{Fig_4Hopping}c, f], since both the characteristic peak in $S(\mathbf{q})$ and well-defined Fermi surface in $n(\mathbf{k})$ disappear. 
{These numerical results suggest the microscopic hopping process $H_\mathrm{QQ}$, 
as illustrated in Fig.~\ref{Fig_Cnnsum}c, is necessary in driving ferromagnetism at finite $U$.}
We further confirm this conclusion by examining a $t_\mathrm{QQ}-U$ model, 
where only $t_\mathrm{QQ}=t$ and finite $U$ are retained, while $t_\mathrm{PQ}=t_\mathrm{QP}=t_\mathrm{PP}=0$.
In this scenario, the itinerant ferromagnetism is observed (see details in Supplementary Note 3).

\begin{figure}[tbp]
\begin{center}
\includegraphics[width=0.49\textwidth]{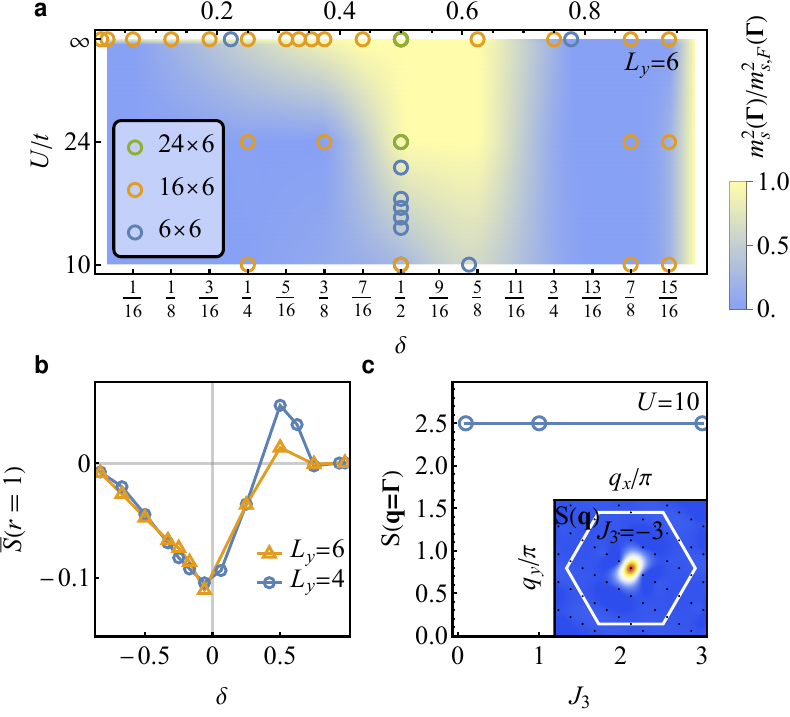}
\end{center}
\par
\caption{
\textbf{$\vert$ Magnetic phase diagram and stability.}
(\textbf{a}) Renormalized ferromagnetic order parameter $m_s^2(\mathbf{\Gamma})/m_{s,\mathrm{F}}^2(\mathbf{\Gamma})$ for the Hubbard model \eqref{eq:model} as a function of electron doping  $\delta$ and $U/t$, {where $U$ is the on-site Hubbard interaction, and $t$ is the nearest-neighbor (NN) hopping amplitude}. Here, $m_{s,\mathrm{F}}^2(\mathbf{\Gamma})\equiv[(1-\delta)/2]^2$ is full polarization value of $m_s^2(\mathbf{\Gamma})$. Circles denote the data obtained numerically.  
(\textbf{b}) Average NN spin correlations $\bar{S}(r=1)$ as a function of doping $\delta$ on cylinders of width $L_y=4,6$, resembling cold-atom quantum simulator findings~\cite{MuqingXu2023} for the same model.
 {
(\textbf{c}) Static structure factor $S(\mathbf{q})$  at the $\boldsymbol{\Gamma}$ point as a function of $J_3$, for the Hubbard model with additional term $\delta H =-J_{3}\sum_{\langle\mathbf{i}\mathbf{j}\rangle} S_{\mathbf{i}}^{z}S_{\text{\ensuremath{\mathbf{j}}}}^{z}$ at $U=10$ {and $8\times4$ cylinder}. The ferromagnetism persists for $J_3 > 0$. Inset:  $S(\mathbf{q})$ for $J_3=3$.
}
}
\label{Fig_PhaseDiagram}
\end{figure}

~\\
\textbf{Magnetic Phase Diagram.}
Now we explore magnetism with a broader doping range and examine both the local and global spin correlations.

The average local spin correlations are defined as 
$
    \bar{S}(r)= \frac{1}{z}\sum_{\mathbf{j},|\mathbf{i}_0-\mathbf{j}|=r}\langle \mathbf{S}_{\mathbf{i}_0}\cdot \mathbf{S}_{\mathbf{j}}\rangle,
$
with the summation over neighbors at a distance $r$ from a fixed site $\mathbf{i}_0$ that is strategically positioned at the lattice's center to avoid a boundary effect. Here, we consider $r=1$ for NN spin correlations, and $z=6$ represents the coordination number of the triangular lattice. 
As shown in Fig.~\ref{Fig_PhaseDiagram}b, the average NN correlations are antiferromagnetic across a broad range of hole doping ($\delta<0$) at intermediate coupling strength $U/t=10$, yet they are markedly suppressed by electron doping ($\delta>0$). In particular, we observe that the local spin correlations become ferromagnetic at intermediate electron doping. 
Notably, these local spin correlations at zero temperature align well with the recent cold-atom experimental findings (figure~3b of Ref.~\citep{MuqingXu2023}) 
at finite temperature for the similar intermediate $U/t$. This observation suggests that the local magnetic properties of the ground state survive in a finite range of temperatures.

We further examine the static spin structure factors to establish the magnetic phase diagram as a function of interaction $U/t$ and electron doping $\delta>0$. 
As illustrated in Fig.~\ref{Fig_PhaseDiagram}a for $L_y=6$ cylinders, the renormalized squared order parameter $m_s^2(\boldsymbol{\Gamma})/m_{s,F}^2(\boldsymbol{\Gamma})$ shows ferromagnetism in a finite range of dopings $\delta \approx 1/2$ and interactions $U\gtrsim10$, which smoothly extends to the infinite interaction limit. These findings are consistent with $L_y=4$ cylinders (see Supplementary Note 4). Together with the numerical evidence presented in the previous section, this demonstrates that the ferromagnetism at both finite and infinite $U$ around $\delta\approx 1/2$ is driven by the doublon-singlon exchange.

Moreover, for finite $U$ and electron doping both below and above $\delta\approx1/2$, the iSDW emerges (see Supplementary Note 4) for both $L_y=4$ and $L_y=6$, which separates the ferromagnetic phase at $n=3/2$ from the phases near the full-filling and half-filling limits. We remark that, unlike the consistent observations at finite coupling strength, at $U=\infty$, we find a smooth connection between the ferromagnetic phase at $\delta \gtrsim 0$ and  $\delta\approx 1/2$ when $L_y=6$, but these two phases are separated by an iSDW phase when $L_y=4$ (see Supplementary Note 4). In addition, for larger doping,  both $L_y=4$ and $L_y=6$ systems exhibit another iSDW phase (see Supplementary Note 4), separating the ferromagnetic regions observed at $\delta\approx1/2$ from ferromagnetism at $\delta\to1$, thus suggesting distinct underlying mechanisms in these two doping concentrations.

~\\
\textbf{Stable Ferromagnetism: Magnetic Anisotropy.}
The ferromagnetic ground state of the Hubbard model spontaneously breaks SU(2) symmetry. Given that stable ferromagnetism is crucial for the development of spintronic devices based on high-temperature ferromagnetic materials, as reported in recent experiments,
we create a spin-wave excitation gap by introducing magnetic anisotropy to stabilize this state.  
Specifically, we consider the following anisotropic term added to the Hubbard Hamiltonian $\delta H =-J_{3}\sum_{\langle\mathbf{i}\mathbf{j}\rangle} S_{\mathbf{i}}^{z}S_{\text{\ensuremath{\mathbf{j}}}}^{z}$,
which can be readily realized by experiments. Here we require $J_{3}>0$.
Using the random phase approximation (RPA) on the Hartree-Fock ground state, we show that the anisotropic term $\delta H$ introduces a gap in the excitation spectrum (see details in Supplementary Note 5), which secures the ferromagnetism even at finite temperatures.

We illustrate the stability of FM phase based on the Hartree-Fock (HF) approximation. We start with the HF Hamiltonian for a FM state with magnetism $m$
\begin{align}\label{eqS:HF}
H_\mathrm{HF} & =\sum_{\mathbf{k},\sigma} (\epsilon_{\mathbf{k}}-\mu) c_{\mathbf{k},\sigma}^{\dagger} c_{\mathbf{k},\sigma} -\frac{1}{2}Um\sum_{\mathbf{k}} c_{\mathbf{k},\alpha}^{\dagger} {\sigma_{\alpha\beta}^{z}} c_{\mathbf{k},\beta} \notag \\
 & =\sum_{\mathbf{k},\sigma}(\epsilon_{\mathbf{k}}-\mu-\frac{1}{2}Um\sigma) c_{\mathbf{k},\sigma}^{\dagger}c_{\mathbf{k},\sigma} \notag \\
 & \equiv\sum_{\mathbf{k}\sigma}\xi_{\mathbf{k},\sigma}c_{\mathbf{k},\sigma}^{\dagger}c_{\mathbf{k},\sigma},
\end{align}
where $\sigma=(\sigma^{x},\sigma^{y},\sigma^{z})$ is the Pauli matrix. We denote the FM ground state of the HF Hamiltonian as $\vert\Omega\rangle$.  Using the random phase approximation (RPA) on the Hartree-Fock ground state, we could derive the spin susceptibility
\begin{equation}
\chi_\mathrm{RPA}^{ij}(\mathbf{q},\omega) = - \frac{\chi_{0}^{+-}(\mathbf{q},\omega)}{\frac{\omega}{mU}-\frac{q^{2}}{2m_{e}mU}}\delta_{ij},\quad i,j=x,y,
\end{equation}
with $m_e$ being the effective mass of electrons.
Here $\chi_0^{+-}(\mathbf q,\omega )$ is the bare spin susceptibility at $xx$, $yy$ directions [see Supplementary Note 5]. 
The susceptibility indicates a spin wave of the square dispersion $\omega(\mathbf q) \propto q^2$. We can decompose the perturbation term $\delta H$, which gives rise to the extra contribution $\delta H_\mathrm{HF}$ to the Hartree-Fock Hamiltonian 
\begin{equation}
\delta H_\mathrm{HF} = - 2zmJ_{3}\sum_{\mathbf{i}}S_{\mathbf{i}}^{z} + zm^{2}J_{3}.
\end{equation}
Here, $z$ is the coordinate number. 
Then the RPA spin susceptibility accordingly changes to 
\begin{equation}
\chi_\mathrm{RPA}^{ij}(\mathbf{q},\omega)=-\frac{\chi_{0}^{+-}(\mathbf{q},\omega)}{\frac{\omega}{mU}-\frac{q^{2}/(2m_e)+2zJ_{3}}{mU} }\delta_{ij},\quad i,j=x,y.
\end{equation}
Obviously, a finite gap $\Delta_\mathrm{gap}=2zJ_3$ is introduced in the spin waves, which can suppress the thermal fluctuations of the spin wave as such to stabilize a FM at finite temperature. We further validate the stability of ferromagnetism under the influence of the anisotropic term $\delta H$ through DMRG simulations across a range of $J_3$ values. As illustrated in Fig.~\ref{Fig_PhaseDiagram}c, we analyze the static structure factor at the $\boldsymbol{\Gamma}$ point as a function of $J_3$.
Our observations indicate that ferromagnetism remains robust for $J_3>0$. Together with the presence of the gap in the dispersion, the DMRG results suggest that ferromagnetism remains stable after the introduction of anisotropy.

~ \\
\textsf{\textbf{\large Discussion}}\\
In this work, we investigate the ferromagnetism mechanism of the finite-doped triangular-lattice Hubbard model with intermediate on-site Coulomb interaction, identifying the cooperative interplay between the charge kinetic process and lattice geometry as the driving factors of ferromagnetism. {By separately diagnosing each of the four microscopic hopping processes decomposed from Hubbard model}, we reveal that the geometrically frustrated doublon-singlon exchange is the only hopping process in triangular lattices for achieving the fully polarized ferromagnetism. We also establish the magnetic phase diagram for the electron-doped triangular Hubbard model and compare the local spin correlations with the latest experiments. urthermore, we demonstrate that the ferromagnetic phase remains robust even when explicit $SU(2)$ symmetry breaking is introduced via anisotropy.

Our findings may serve as the starting point and stimulate future studies closely related to distinct TMD hetero- and homo-bilayer materials, such as exploring the stability of ferromagnetism against long-range Coulomb interactions and complex hoppings. Notably,  an insulating stripe phase at the $\delta=1/2$ electron doping was recently found experimentally on transition-metal dichalcogenides (TMD) moir\'{e} superlattices formed by hetero-bilayers of WSe$_2$/WS$_2$  \cite{JinChenhao2021}, which are believed to be described by the extended Hubbard models in the strong coupling limit with long-range Coulomb interactions $V$. This revelation invites further inquiries into the stability of ferromagnetism under the influence of long-range Coulomb interactions, the emergence of other correlated phases \cite{XuYang2020}, and the possible direct quantum phase transitions among these conventional symmetry-breaking phases \cite{Senthil2004Science,Senthil2004,Senthil2023}.
In addition, another type of TMD moir\'{e} superlattices formed by homo-bilayers \cite{LeiWang2020} has been proposed to be described by the moir\'{e} Hubbard model, which generalizes the pure Hubbard model by introducing complex phases in hoppings $te^{i\sigma\phi}$. Therefore, our work may further motivate
the investigations into how the hopping phase influences the emergence of kinetic-energy-driven ferromagnetism. Moreover, these findings may also shed light on the magnetism with general interaction in diverse lattice geometries~\cite{Mielke1991, Mielke1992,Wurth1996,ChiaChenChang2010,Scholle2023,Riegler2023, Scholle2024}.

\begin{figure}[t]
\begin{center}
\includegraphics[width=0.48\textwidth] {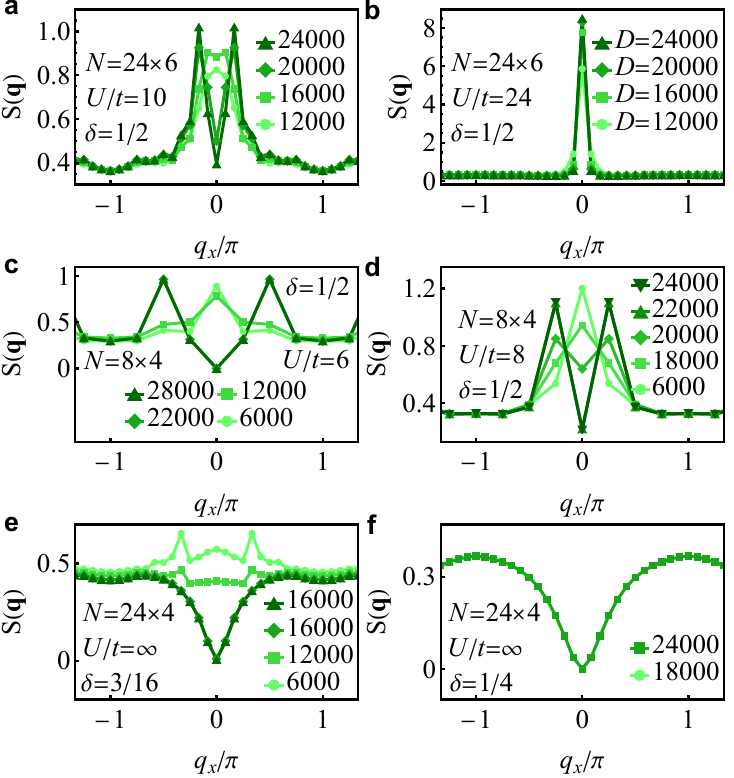}
\end{center}
\par
\caption{
\textbf{$\vert$ Line-cut plot of the static spin structure $\boldsymbol{ S(\mathbf{q})}$ with respect to various bond dimensions $\boldsymbol D$.} The line-cut path is chosen along the momentum path passing through the $\Gamma$ point (denoted as the white dashed line in Fig.~\ref{Fig_FMPropertySq}b in the main text). 
{Parameters for system size $N$, coupling $U/t$, electron doping $\delta$: 
(\textbf{a}) $N=24\times 6$, $U/t=10$, $\delta=1/2$; 
(\textbf{b}) $N=24\times 6$, $U/t=24$, $\delta=1/2$; 
(\textbf{c}) $N=8\times 4$, $U/t=6$, $\delta=1/2$; 
(\textbf{d}) $N=8\times 4$, $U/t=8$, $\delta=1/2$; 
(\textbf{e}) $N=24\times 4$, $U/t=\infty$, $\delta=3/16$; 
(\textbf{f}) $N=24\times 4$, $U/t=\infty$, $\delta=1/4$.
}
}
\label{FigS_UBonddimQxatGammaSqDelta}
\end{figure}
~ \\
\textsf{\textbf{\large METHODS}}
~\\
\noindent\textbf{Density matrix renormalization group} \\
In the numerical part, we employ DMRG algorithm to simulate the Hubbard model Hamiltonian in Eq.~\eqref{eq:model} on triangular lattice cylinders with primitive vectors $\boldsymbol{e}_x=(1,0)$, $\boldsymbol{e}_y=(1/2,\sqrt{3}/2)$, wrapping on cylinders with a lattice spacing of unity. System size is $N=L_x\times L_y$, where $L_x$ ($L_y$) represents the length (circumference) of the cylinder. Here we mainly study cylinders with width $L_y=4$, $6$ and set the energy unit as $t$. {We remark that, for the triangular lattice, $\mathbf{K}$ points are inaccessible on cylinders with width $L_y=4$. 
As a result,  it is hard to exactly capture the magnetic orders with periodicity of 3, such as the $120^{\circ}$ N\'eel order. Therefore, we also examine the $L_y=6$ cylinders to suppress possible finite-size effects.
}

Our numerical simulation is based on the Hubbard model and our focus is the intermediate $U$ Hubbard model. Only in the infinite-$U$ limit, we use the effective $t$-$J$ model with vanishing superexchange interaction $J=4t^2/U$. 

For DMRG calculations, given the varying convergence rates at different parameters, we set the DMRG bond dimension $D$ up to $D\approx 36,000$.  Notably, large bond dimensions are necessary for distinguishing between ferromagnetic and incommensurate spin-density-wave (SDW) phases.  
We provide specific examples to illustrate the necessity of large bond dimensions for accurately identifying the ground-state properties of both the incommensurate SDW and ferromagnetic phases. Additionally, we have performed DMRG calculations with different initial states to ensure the convergence of DMRG. (1) For incommensurate SDW region, at relatively small bond dimensions, the peak of the static spin structure $S(\mathbf{q})$ is located at the momentum $\mathbf{\Gamma}$ [see Figs.~\ref{FigS_UBonddimQxatGammaSqDelta}a, c, d], which is a signature for ferromagnetism. However, with the increase of bond dimension, the energy becomes lower and the peak splits around the momentum $\mathbf{\Gamma}$. 
These observations indicate large bond dimensions are required to obtain converged results for distinguishing the ferromagnetism and iSDW with close energies. (2) On the other side, for the ferromagnetism region, a large bond dimension is crucial for determining whether the ground state is partially polarized or fully polarized. As presented in Fig.~\ref{FigS_UBonddimQxatGammaSqDelta}b, the peak of $S(\mathbf{q})$ gets higher with increasing bond dimension, until it saturates. Notably, DMRG convergence becomes more challenging with decreasing interaction strength $U/t$ and increasing system size $N$. In Figs.~\ref{FigS_UBonddimQxatGammaSqDelta}e, f, we also present the good convergence of the DMRG simulations in the iSDW region at $U=\infty$ that separates the ferromagnetic phases at $\delta\approx1/16$ and $\delta\approx1/2$ for $L_y=4$ [see 
Supplementary Note 4].
\\

\noindent\textbf{Unrestricted Hartree-Fock analysis}\\
In the analytical part, we apply the unrestricted Hartree-Fock (UHF)
approximation. The mean-field order parameters
are introduced for both the on-site densities $\langle n_{\mathbf i,\sigma}\rangle$
and the spin-flips $\langle S_{\mathbf i}^{-}\rangle$ and $\langle S_{\mathbf i}^{+}\rangle$. We have a total $4N$ variational parameters with $N$ being the number of the lattice sites. These variational parameters can be iteratively solved by the
self-consistent equations
\begin{align}
\langle n_{\mathbf i,\sigma}\rangle & =\sum_{n}f_{FD}(E_{n})\langle\psi_{n}\vert n_{\mathbf i,\sigma}\vert\psi_{n}\rangle,\\
\langle S_{\mathbf i}^{\pm}\rangle & =\sum_{n}f_{FD}(E_{n})\langle\psi_{n}\vert S_{\mathbf i}^{\pm}\vert\psi_{n}\rangle,
\end{align}
where $\vert\psi_{n}\rangle$ is an eigenstate of $H_\mathrm{HF}$ with $H_\mathrm{HF}\vert\psi_{n}\rangle=E_{n}\vert\psi_{n}\rangle$
and $f_{FD}(E_{n})$ is the Fermi-Dirac distribution.

For a stable convergence, we utilize the direct inversion in
the iterative subspace (DIIS) method. In this approach, we can express
the Hartree-Fock Hamiltonian in the matrix form as follows: 
\begin{equation}
H_\mathrm{HF}=C^{\dagger}FC+\sum_{\mathbf i}U\left[-\langle n_{\mathbf i,\uparrow}\rangle\langle n_{\mathbf i,\downarrow}\rangle+\langle S_{\mathbf i}^{-}\rangle\langle S_{\mathbf i}^{+}\rangle\right],
\end{equation}
where $C$ is a vector with $C_{\mathbf i\sigma}=c_{\mathbf i,\sigma}$ and $F$
is the Fock matrix. For a magnetic order, the single-particle correlation
matrix $\Phi$ is defined as 
$
\Phi_{\mathbf{ij},\sigma\sigma^\prime}=\langle c_{\mathbf i,\sigma}^{\dagger}c_{\mathbf j,\sigma^\prime}^{}\rangle.
$
The DIIS method relies on the property that a convergence condition
is met when the single-particle correlation matrix commutes with the
Fock matrix
$
F\Phi-\Phi F=0.
$
This condition allows us to introduce an error vector for the $n$-th iteration
as 
$
\mathbf{e}_{ n}=F_{ n}\Phi_{n}-\Phi_{n}F_{n},
$
where $\Phi_{n}$ is obtained from the diagonalization of $F_{n}$. The
iteration process is to  update the Fock matrix $F_{n}$ at $n$th step as 
$
F_{n}=\sum_{i=1}^{k}d_{n-i}F_{n-i},
$
by involving  the last $k$ Fock matrices ${F_{n-i}}$ ($i=1,\cdots,k$).
Here, $d_{n-i}$ are the DIIS coefficients obtained through a least-squares
constrained minimization of the error vectors. That is, we perform the
optimization by minimizing the error
$
\epsilon  =\vert\sum_{i=1}^{k}d_{n-i}\mathbf{e}_{n-i}\vert^{2}
$
under the constraint $\sum_{i=1}^{k}d_{n-i}=1$. In our calculations, we set the convergent condition of the energy difference between two subsequent iteration processes as $10^{-8}t$.

~\\~
\vspace{0.5cm}

 ~ \\
 \textsf{\textbf{\large DATA AVAILABILITY}}\\
The data that support the findings of this study are available from the corresponding author upon reasonable request.

{
 ~ \\
 \textsf{\textbf{\large CODE AVAILABILITY}}\\
All numerical codes in this paper are available upon request to the authors.
}

~ \\
{\textbf{\large Acknowledgements}} \\
We thank helpful discussions with Ashvin Vishwanath and Yuchi He.
This work is supported by the National Natural Science Foundation of China (Grant No.92477106) and the Fundamental Research Funds for the Central Universities.

~ \\
{\textbf{\large Author contributions}} \\
All authors contributed to developing the ideas, analysing the results and writing the manuscript. Z.Z. initiated and led the project. Q.Q.C. performed the numerical calculations and data analysis of the main results. S.A.C. performed the mean-field analysis.

~ \\
{\textbf{\large Competing interests}} \\
The authors declare no competing interests.

\onecolumngrid
\newpage
\renewcommand{\theequation}{S\arabic{equation}}
\setcounter{equation}{0}
\renewcommand{\thefigure}{S\arabic{figure}}
\setcounter{figure}{0}
\renewcommand{\bibnumfmt}[1]{[S#1]}
\begin{center}
  \textbf{Supplementary Information for\\``{Geometric Frustration Assisted Kinetic Ferromagnetism in Doped Mott Insulators}"}
\end{center}

\author{Qianqian Chen}
\affiliation{Kavli Institute for Theoretical Sciences, University of Chinese Academy of Sciences, Beijing 100190, China}
\author{Shuai A. Chen}
\affiliation{Max Planck Institute for the Physics of Complex Systems, N\"{o}thnitzer Stra{\ss}e 38, Dresden 01187, Germany}
\author{Zheng Zhu}
\affiliation{Kavli Institute for Theoretical Sciences, University of Chinese Academy of Sciences, Beijing 100190, China}

\maketitle

\section{Evolution of the magnetic structure factor at Electron Doping $\delta=1/2$} \label{sec:AppendixEvolutionOfQ0}

\begin{figure}[b]
\begin{center}
\includegraphics[width=0.45\textwidth] {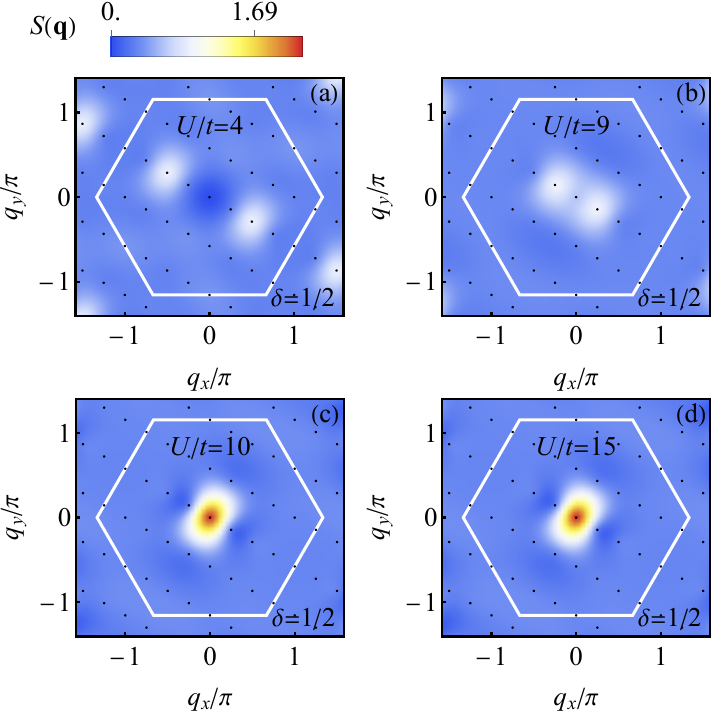}
\end{center}
\par
\renewcommand{\figurename}{Fig.}
\caption{
\textbf{The contour plot of the static spin structure factor $\boldsymbol{ S(\mathbf{q})}$.} Calculated with $N=8\times 4$ at $\delta=1/2$ for the Hubbard model.
}
\label{FigS_UQxQySqDelta05N8x4}
\end{figure}

\begin{figure}[tb]
\begin{center}
\includegraphics[width=0.45\textwidth]{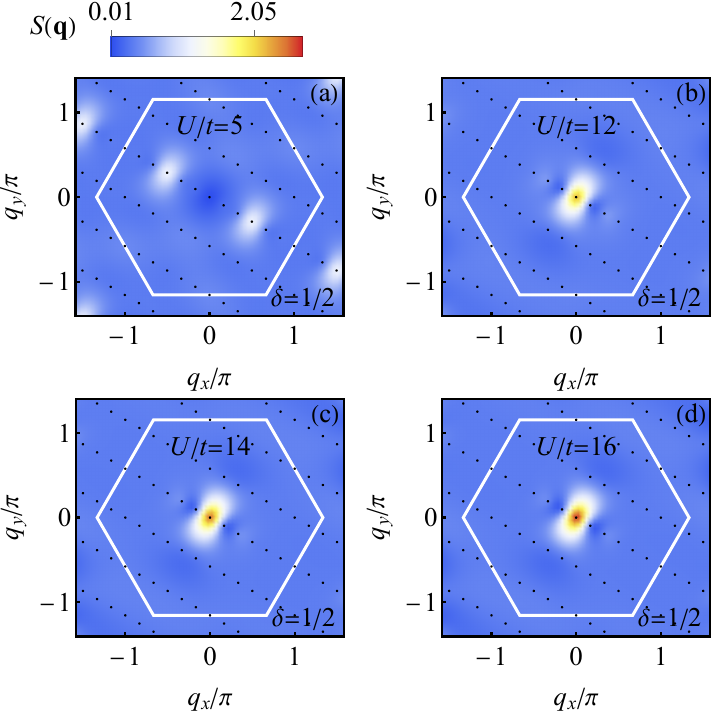}
\end{center}
\par
\caption{
\textbf{The contour plot of the static spin structure factor $\boldsymbol{ S(\mathbf{q})}$.} Calculated with $N=12\times 4$ at $\delta=1/2$ for the Hubbard model.
}
\label{FigS_UQxQySqDelta05N12x4}
\end{figure}

We present additional supporting data for the evolution of the momentum $\mathbf{q}_0$ of the peak in $S(\mathbf{q})$ as a function of interaction strength $U/t$ in  Fig.~\ref{FigS_UQxQySqDelta05N8x4} for $N=8\times 4$ and
Fig.~\ref{FigS_UQxQySqDelta05N12x4} for $N=12\times 4$. These results for $L_y=4$ systems are consistent with the case of $L_y=6$ in Figs.~1(a)-(b) in the main text. 
The peak locations in $S(\mathbf{q})$ progressively shift toward the $\mathbf{\Gamma}$ point as $U/t$ increases, as presented in Fig.~\ref{FigS_UQxQySqDelta05N8x4}(a)-(c) and Figs.~\ref{FigS_UQxQySqDelta05N12x4}(a), (b). This observation supports the schematic plot in Fig.~1(c) in the main text. 
After the peak of $S(\mathbf{q})$ shifts to the $\mathbf{\Gamma}$ point, its height increases within a specific range of $U/t$ [see Figs.~\ref{FigS_UQxQySqDelta05N12x4}(b)-(d)], until reaching its maximum value, as illustrated by Figs.~\ref{FigS_UQxQySqDelta05N8x4}(c), (d),  and
Figs.~\ref{FigS_UQxQySqDelta05N12x4}(c), (d), indicating a transition from a partially-polarized to a fully-polarized ferromagnetic phase.

\section{{Perturbation and the wave function analysis for ferromagnetic mechanism}}\label{sec:AppendixPertWaveF}
As shown in the main text, the Hubbard Hamiltonian encompasses several types of hopping processes.
Our measurements of each microscopic process reveal that the ferromagnetism at finite $U$ arises solely from the doublon-singlon exchange. 
Therefore, in the following, we concentrate on the doublon-singlon exchange, and aim to understand why ferromagnetism is present on the electron-doped cases but absent on the hole-doped cases. To address this, we have developed two lines of reasoning, i.e., the perturbation analysis and the wave function analysis.

\begin{figure}[hb]
\begin{center}
\includegraphics[width=0.48\textwidth]{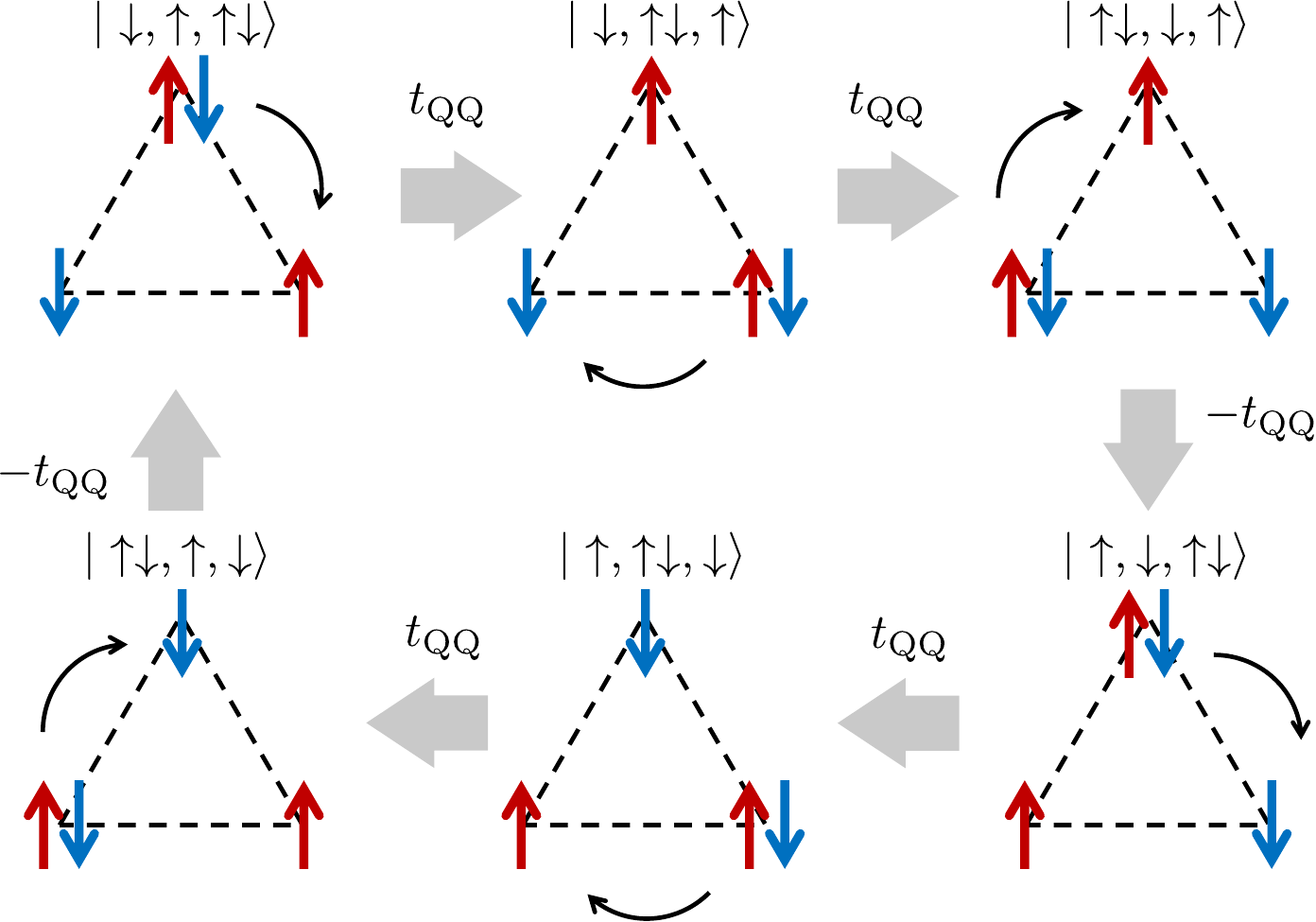}
\end{center}
\par
\renewcommand{\figurename}{Fig.}
\caption{
\textbf{Schematic illustration of hopping processes in the electron-doped Hubbard model. }
Here, we have taken the denotation of the bases into consideration for the hopping signs.
}
\label{FigS_hoppingAmp_ElectronDope}
\end{figure}

For the perturbation analysis, we consider two types of effective spin exchange resulting from kinetic virtual processes: Superexchange, commonly encountered in the Hubbard model, giving rise to the term $\sim {t^2}/{U}\mathbf{S}_i\cdot\mathbf{S}_j$; and doublon-assisted ring exchange, represented as $\sim t^3/\Delta E^2 \mathbf{S}_i\cdot\mathbf{S}_j$, 
where $\Delta E$ is the energy difference between the initial and intermediate states.
In the ring exchange process, a doubly occupied site is necessary for one electron to exchange while another passes by, as illustrated in Fig.~3(d) in the main text. 
Notably, $\Delta E$ is of the order of $t$ because the presence of two neighboring doublons restricts the hopping of singlons, preventing them from gaining kinetic energy.
{For electron doping (where $t<0$), the virtual processes involved in ring exchange $t^3/\Delta E^2$ promote ferromagnetism and dominates over the superexchange coupling $t^2/U$ for $U>t$, since from $\Delta E\sim |t|<{U} $ we have $|t^3/\Delta E^2|>t^2/U$}. On the other side, for hole doping (where $t>0$), the ring exchange $t^3/\Delta E^2$ favor antiferromagnetism, similar to the effect of superexchange coupling.

\begin{figure}[htb]
\begin{center}
\includegraphics[width=0.465\textwidth]{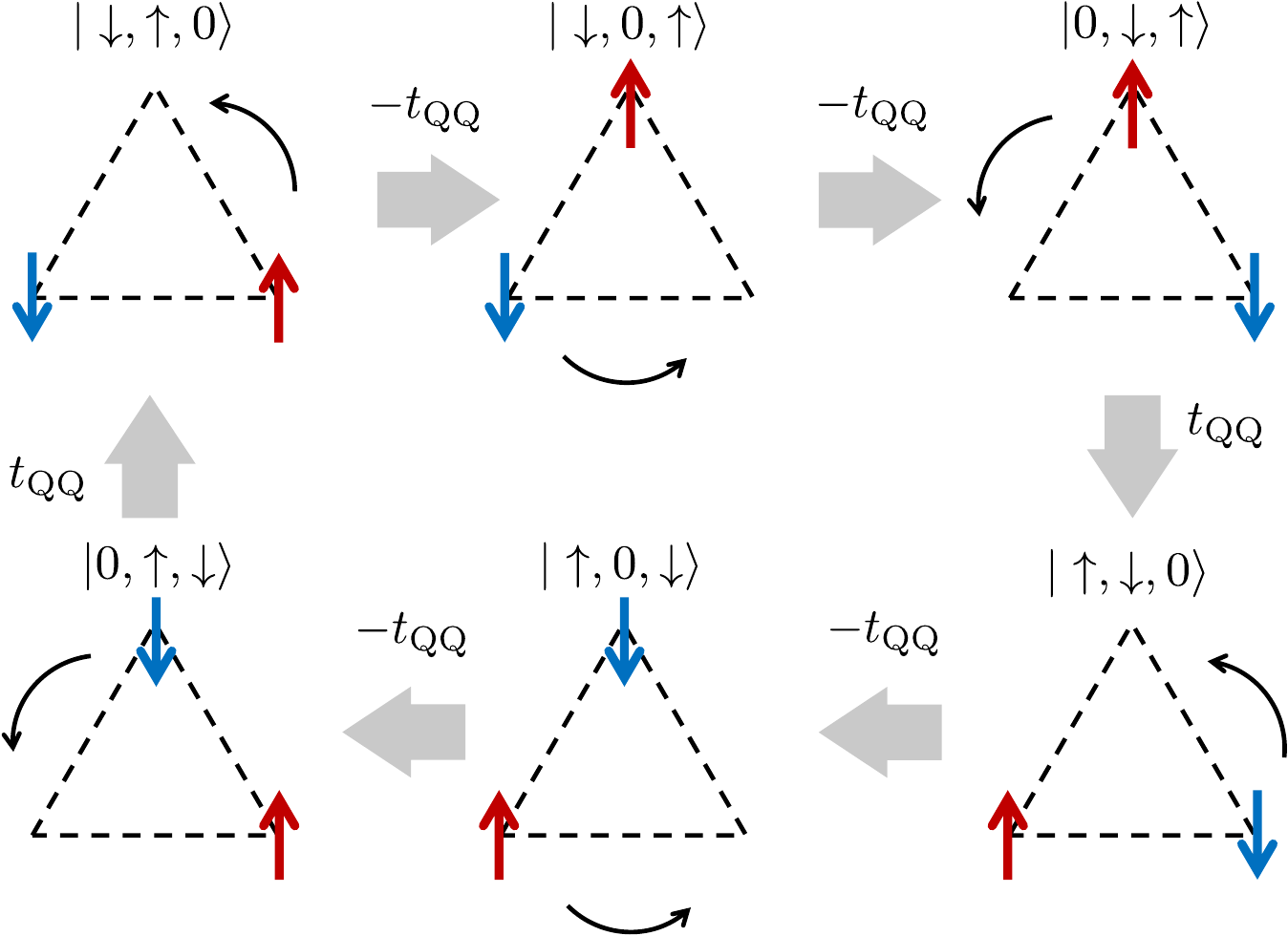}
\end{center}
\renewcommand{\figurename}{Fig.}
\caption{
\textbf{Schematic illustration of hopping processes in the hole-doped Hubbard model. }
Here, we have taken the denotation of the bases into consideration for the hopping signs.
}
\label{FigS_hoppingAmp_holeDope}
\end{figure}

\begin{figure}[b]
\begin{center}
\includegraphics[width=0.48 \textwidth]{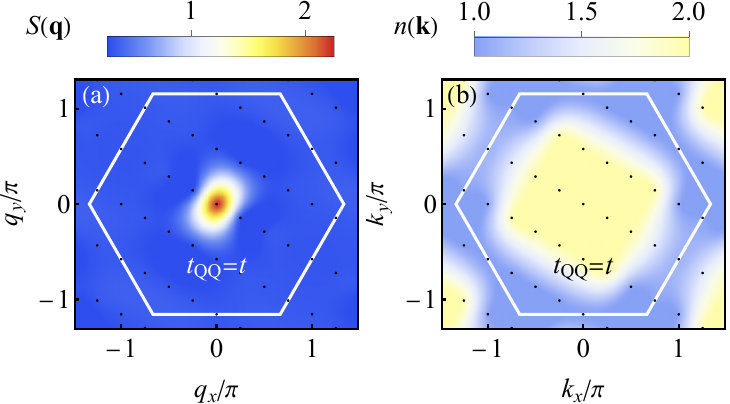}
\end{center}
\par
\renewcommand{\figurename}{Fig.}
\caption{
\textbf{Itinerant ferromagnetism signature for the $\boldsymbol{ t_\mathrm{QQ}-U}$ model.}
(\textbf{a}) The static spin structure factor $S(\mathbf{q})$. 
(\textbf{b}) The electron momentum distribution $n(\mathbf{k})$.
The black dots represent the accessible momenta in the Brillouin zone (white lines). Interpolation has been applied in the contour plot. Here, $N=8\times 4$, $U/t=10$, $\delta=1/2$.
}
\label{FigS_T1ET2ET3E0U10N8x4Delta05}
\end{figure}

We also offer an alternative wave function analysis in the large $U$ limit of the Hubbard model with $t>0$.
We focus on $S_z=0$ sector because the ground state resides in this sector regardless of whether the system exhibits ferromagnetism.
The relevant basis states and the transitions between them are depicted in Fig.~\ref{FigS_hoppingAmp_ElectronDope}.
In Fig.~\ref{FigS_hoppingAmp_ElectronDope}, we carefully account for the signs of the hopping matrix element by considering the construction of the basis, i.e., the requirement of the normal order for the creation operators acting on the vacuum state. For example: 
\begin{equation}
    |\uparrow, \downarrow, \uparrow \downarrow\rangle=c_{1, \uparrow}^{\dagger} c_{2, \downarrow}^{\dagger} c_{3, \uparrow}^{\dagger} c_{3, \downarrow}^{\dagger}|0\rangle.
\end{equation}
To determine the ground state $\left|\Psi_0^e\right\rangle$ for the electron-doped case, we impose the system's symmetry and minimize the energy by choosing appropriate superpositions of basis states. The signs of the coefficients are determined based on the following principle: If the hopping matrix element connecting two configurations is negative (positive), superposing these configurations with the same (opposite) sign leads to the minimization of the energy.
With these considerations, therefore, the ground state must have the following form
\begin{equation}
\begin{aligned}
|\Psi_0^e\rangle\propto&\frac{\sqrt{2}}{2}\left( |\uparrow,\downarrow,\uparrow\downarrow\rangle+|\downarrow,\uparrow,\uparrow\downarrow\rangle
\right)\\
&-\frac{\sqrt{2}}{2}\left(
|\uparrow,\uparrow\downarrow,\downarrow\rangle+|\downarrow,\uparrow\downarrow,\uparrow\rangle
\right)\\
&
+\frac{\sqrt{2}}{2}\left(
|\uparrow\downarrow,\uparrow,\downarrow\rangle+|\uparrow\downarrow,\downarrow,\uparrow\rangle
\right).
\end{aligned}
\end{equation}
The state $\left|\Psi_0^e\right\rangle$ is symmetric under the exchange of spins and sites, thus corresponds to a triplet state, indicating ferromagnetic ordering. 
Conversely, for the hole-doped case, the transitions between the basis states are illustrated in Fig.~\ref{FigS_hoppingAmp_holeDope}. Following the same methodology, the ground state $|\Psi_0^h\rangle$ is constructed as:
\begin{equation}
\begin{aligned}
|\Psi_0^h\rangle\propto&\frac{\sqrt{2}}{2}\left( |\uparrow,\downarrow,0\rangle-|\downarrow,\uparrow,0\rangle
\right)\\
&
+\frac{\sqrt{2}}{2}\left(
|\uparrow,0,\downarrow\rangle-|\downarrow,0,\uparrow\rangle
\right)\\
&
+\frac{\sqrt{2}}{2}\left(
|0,\uparrow,\downarrow\rangle-|0,\downarrow,\uparrow\rangle
\right).
\end{aligned}
\end{equation}
In this case, the coefficients are chosen to produce antisymmetric combinations under the exchange of spins, resulting in a singlet state with total spin $S=0$.

\begin{figure}[b]
\begin{center}
\includegraphics[width=0.5 \textwidth]{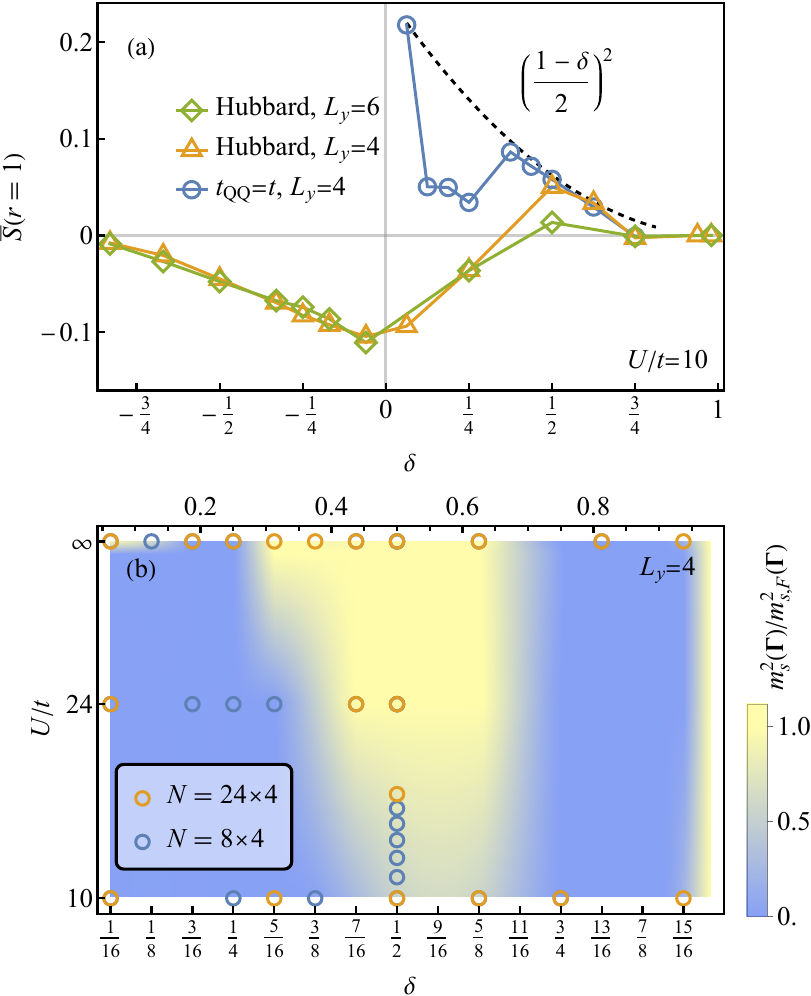}
\end{center}
\par
\renewcommand{\figurename}{Fig.}
\caption{
(\textbf{a}) The average spin correlations $\bar{S}(r=1)$, as defined in the main text and indicated by yellow triangles ($L_y=4$) and green diamonds ($L_y=6$), are calculated across the nearest-neighbor bonds for the standard Hubbard model.
These results are compared with those (blue circles) obtained when $t_\mathrm{QQ}=t$ in Eq.~(6) in the main text, with all other hopping processes absent. The dashed black line signifies the fully polarized case. The largest system sizes we presented here are $N=16\times 6$ and $N=18\times 4$ for 6- and 4-leg cylinders, respectively.  
(\textbf{b}) The renormalized ferromagnetic order parameter $m_s^2(\mathbf{\Gamma})/m_{s,\mathrm{F}}^2(\mathbf{\Gamma})$ for the Hubbard model as a function of doping  $\delta$ and interaction $U/t$. Here, $m_{s,\mathrm{F}}^2(\mathbf{\Gamma})\equiv[(1-\delta)/2]^2$  corresponds to a full polarization value of $m_s^2(\mathbf{\Gamma})$. Circles denote the data obtained numerically. 
}
\label{FigS_DeltaSrnnU10ANDUDeltaSGammaLy4}
\end{figure}

\section{{{Itinerant Ferromagnetism  of $t_{\mathrm{QQ}}-U$ model}}}\label{sec:AppendixTQQU}
The $t_{\mathrm{QQ}}-U$ model is given by $H=H_{t_\mathrm{QQ}}+H_\mathrm{int}$, where the interaction term is the Hubbard repulsion $H_\mathrm{int}=U\sum_\mathbf{i} n_{\mathbf{i}\uparrow} n_{\mathbf{i}\downarrow}$, and the projected hopping term is given by 
\begin{equation}\label{eq:tQQ}
    H_{t_\mathrm{QQ}}=-\sum_{\langle \mathbf{i} \mathbf{j}\rangle, \sigma}   t_\mathrm{QQ} \hat{Q}_{\mathbf{i},\bar{\sigma}} c_{\mathbf{i},\sigma}^{\dagger} c_{\mathbf{j},\sigma} \hat{Q}_{\mathbf{j},\bar{\sigma}}.
\end{equation}
Here, the projection operator is $\hat{Q}_{\mathbf{i},{\sigma}}\equiv n_{\mathbf{i}, {\sigma}}$ and Eq.~\eqref{eq:tQQ} describes the NN hoppings between a doubly occupied site and a singly occupied site with hopping amplitude $t_\mathrm{QQ}$ [see Fig.~3(c) in the main text], i.e., the doublon-singlon exchange. In Fig.~\ref{FigS_T1ET2ET3E0U10N8x4Delta05}, we examine both the static spin structure factor $S(\mathbf{q})$ and electron momentum distribution $n(\mathbf{k})$ at electron doping $\delta=1/2$ for $U/t=10$, where the pronounced peak at $\boldsymbol{\Gamma}$ and the well-defined Fermi surface are consistent with the itinerant ferromagnetism observed in the Hubbard model at the same doping and the same coupling strength $U/t=10$.

\section{{Evolution of Magnetism with Doping}}\label{sec:AppendixPhaseDiagram}
In this section, we present the average spin correlations $\bar{S}(r=1)$ and the magnetic phase diagram as a function of $U/t$ and electron doping $\delta$ specifically for $L_y=4$. We also provide further data for $L_y=6$ to enhance the comparison.

\begin{figure}[b]
\begin{center}
\includegraphics[width=0.48\textwidth]{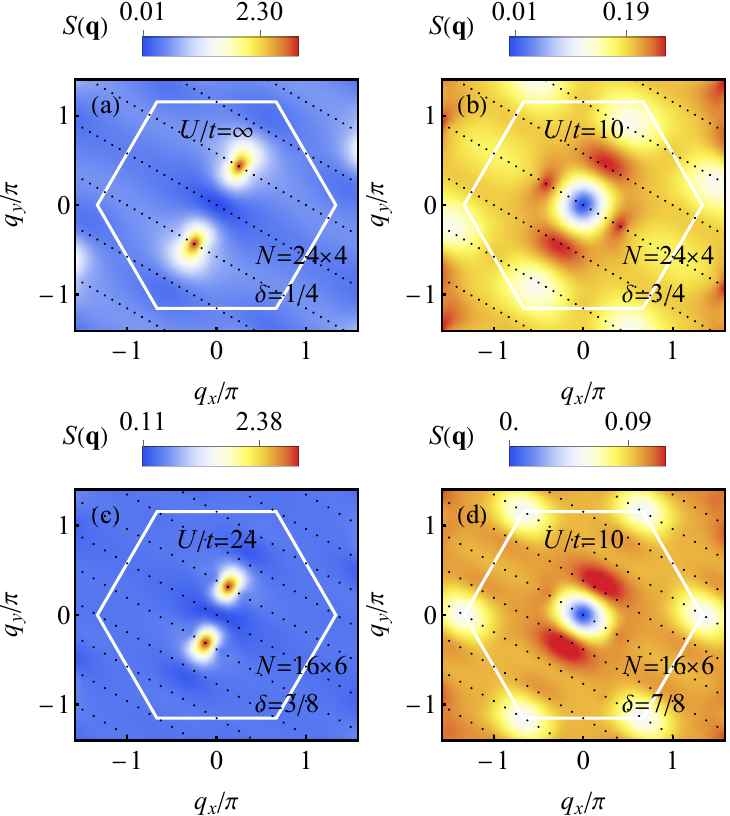}
\end{center}
\par
\renewcommand{\figurename}{Fig.}
\caption{
\textbf{Contour plots of the static spin structure factor $\boldsymbol{ S(\mathbf{q})}$  for the Hubbard model, illustrating two distinct iSDW phases. }
Panels (\textbf{a}) and (\textbf{c}) show phases at smaller doping, while panels (\textbf{b}) and (\textbf{d}) depict phases at larger doping. 
}
\label{FigS_QxQySqiSDW}
\end{figure}

For the fully polarized phase of the Hubbard model with $U/t=10$, $\delta\approx1/2$, $L_y= 4$, the average spin correlations $\bar{S}(r=1)$ [yellow triangles in Fig.~\ref{FigS_DeltaSrnnU10ANDUDeltaSGammaLy4}(a)] align well with those [blue circles] of the $t_\mathrm{QQ}-U$ model Eq.~\eqref{eq:tQQ}. 
This alignment underscores a deep connection between the ferromagnetic mechanisms inherent to both cases, and thus highlights the significant influence of doublon-singlon exchange in inducing ferromagnetism in the Hubbard model with finite $U/t$.  
In addition,  for light electron doping around $\delta\approx1/16$, the  $\bar{S}(r=1)$  of the $t_\mathrm{QQ}-U$ model [blue circles in Fig.~\ref{FigS_DeltaSrnnU10ANDUDeltaSGammaLy4}(a)] 
closely matches with the fully polarized value (dashed black line) $m_{s,\mathrm{F}}^2(\mathbf{\Gamma})\equiv[(1-\delta)/2]^2$, yet deviation becomes apparent around a doping range of  $3/16\lesssim\delta\lesssim1/4$, the same as the ferromagnetic feature of the Hubbard model at $U/t=\infty$ for $L_y=4$, as shown by the Fig.~\ref{FigS_DeltaSrnnU10ANDUDeltaSGammaLy4}(b).

Figure~\ref{FigS_DeltaSrnnU10ANDUDeltaSGammaLy4}(b) presents the magnetic phase diagram as a function of $U/t$ and electron doping $\delta$  for $L_y=4$, identifying two distinct ferromagnetic phases at $U=\infty$ around doping $\delta\approx 1/16$ and $\delta\approx 1/2$, separated by a doping range of  $3/16\lesssim\delta\lesssim1/4$.  The pattern of $S(\mathbf{q})$ at such separation [see Fig.~\ref{FigS_QxQySqiSDW}(a)] is similar to that with finite $U$ [see Fig.~\ref{FigS_QxQySqiSDW}(c)], indicating that the iSDW at small doping extends from finite $U$ to $U=\infty$ smoothly for the case of $L_y=4$.  
This numerical observation suggests the subtle competition between the iSDW and FM at intermediate doping on narrower cylinders like $L_y=4$ and the final stabilization of FM on wider cylinders  like $L_y=6$ [see Fig.~5(a) in the main text]. 
The notable separations of ferromagnetic phases around $\delta\approx 1/16$, $\delta\approx 1/2$ and $\delta\to 1$ for a finite system with $L_y=4$ hint at diverse underlying mechanisms, potentially differentiating the observed ferromagnetism around $\delta\approx 1/2$ from the Nagaoka ferromagnetism with $\delta\to 0$ and the Mielke’s flat-band ferromagnetism or Müller-Hartmann ferromagnetism with $\delta\to 1$ for a finite size system, even though such separations of the ferromagnetic phases are absent for $L_y=6$ [see Fig.~5(a) in the main text] at least for the largest length $L_x=16$ that we have simulated.  
On the other hand, as the effective interaction $U/t$ decreases from $\infty$ to $U/t=10$ for $L_y=4$, the ferromagnetic region around lighter ($\delta\approx1/16$) and intermediate ($\delta\approx 1/2$) doping disappears and shrinks [Fig.~\ref{FigS_DeltaSrnnU10ANDUDeltaSGammaLy4}(b)], respectively, yielding a ferromagnetic landscape closely aligned with the $U/t=10$ case as depicted in Fig.~\ref{FigS_DeltaSrnnU10ANDUDeltaSGammaLy4}(a). 
Additionally, for larger doping $5/8\lesssim\delta< 1$, a distinct pattern in $S(\mathbf{q})$ indicates the presence of another  iSDW phase [see Figs.~\ref{FigS_QxQySqiSDW}(b) and (d)].

We further present the  $S(\mathbf{q})$ outside the ferromagnetism phase at small doping with finite $U$ and larger doping $5/8\lesssim \delta<1$ in Fig.~\ref{FigS_QxQySqiSDW}. Our examination reveals two distinct features in $S(\mathbf{q})$, each indicative of a different iSDW phase.  

\section{{RPA analysis for Stabilization of Ferromagnetism via Magnetic Anisotropy}}\label{sec:AppendixRPA}

In this section, we create a spin-wave excitation gap by introducing magnetic anisotropy into the system, which stabilize the FM. 
Such stabilization mechanisms are relevant not only for fundamental physics but also for the development of spintronic devices based on high-temperature ferromagnetic materials, as reported in recent experiments \cite{ChengGong2017,Bonilla2018, YujunDeng2018,Gibertini2019,ChengGong2019, BedoyaAmilcar2021, HaoWang2022}. 
As an example, we consider the following anisotropic term added to the Hubbard Hamiltonian
\begin{equation}\label{eqS:J3}
\delta H =-J_{3}\sum_{\langle\mathbf{i}\mathbf{j}\rangle} S_{\mathbf{i}}^{z}S_{\text{\ensuremath{\mathbf{j}}}}^{z},
\end{equation}
which can be readily realized by experiments, such as in a multilayer system. Here we require $J_3>0$. The anisotropic term in Eq.~(\ref{eqS:J3}) breaks the spin $SU(2)$ symmetry. In the following, we apply the random phase approximation (RPA) on the HF ground state to show that the anisotropic term \eqref{eqS:J3} can induce a finite gap of the spin-wave spectrum, which then protects the ferromagnetic ground state at finite temperatures. The gap can be expressed as
\begin{equation}\label{eqS:gap}
    \Delta =2z J_3,
\end{equation}
where $z$ is the coordinate number.

\subsection{Spin-wave dispersion without magnetic anisotropy}
We illustrate the instability of FM phase based on the Hartree-Fock (HF) approximation. We start with the HF Hamiltonian for a FM state with magnetism $m$
\begin{align}
H_\mathrm{HF} & =\sum_{\mathbf{k},\sigma}(\epsilon_{\mathbf{k}}-\mu)c_{\mathbf{k},\sigma}^{\dagger} c_{\mathbf{k},\sigma}-\frac{1}{2}Um\sum_{\mathbf{k}} c_{\mathbf{k},\alpha}^{\dagger} {\sigma_{\alpha\beta}^{z}} c_{\mathbf{k},\beta} \notag \\
 & =\sum_{\mathbf{k},\sigma}(\epsilon_{\mathbf{k}}-\mu-\frac{1}{2}Um\sigma) c_{\mathbf{k},\sigma}^{\dagger}c_{\mathbf{k},\sigma} 
 \equiv\sum_{\mathbf{k}\sigma}\xi_{\mathbf{k},\sigma}c_{\mathbf{k},\sigma}^{\dagger}c_{\mathbf{k},\sigma},\notag
\end{align}
where $\sigma=(\sigma^{x},\sigma^{y},\sigma^{z})$ is the Pauli matrix. 
We denote the FM ground state of the HF Hamiltonian as $\vert\Omega\rangle$. 
We can uncover the spin wave excitation by studying the charge and spin susceptibility
\begin{align}
\chi^{00}(\mathbf{q},t) & =\frac{i}{2N}\langle T\rho_{\mathbf{q}}(t)\rho_{-\mathbf{q}}(0)\rangle,\\
\chi^{ij}(\mathbf{q},t) & =\frac{i}{2N}\langle TS_{\mathbf{q}}^{i}(t)S_{-\mathbf{q}}^{j}(0)\rangle,
\end{align}
with the density and spin operators $\rho_{\mathbf{q}}$ and $S_{\mathbf{q}}^{i}$ ($i=x,y,z$)
\begin{align}
\rho_{\mathbf{q}} =\sum_{\mathbf{k},\sigma}c_{\mathbf{k}+\mathbf{q},\sigma}^{\dagger}c_{\mathbf{k},\sigma}^{},\quad
S_{\mathbf{q}}^{i} =\sum_{\mathbf{k},\alpha,\beta}c_{\mathbf{k}+\mathbf{q},\alpha}^{\dagger} c_{\mathbf{k},\beta}\sigma_{\alpha\beta}^{i}.
\end{align} 
For the notation simplicity, we ignore an extra $\frac{1}{2}$ factor in $S_\mathbf{q}^i$.

Regarding the HF Hamiltonian, we can first define the charge and spin susceptibility
\begin{align}
    \chi_{0}^{00}(\mathbf{q},\omega) & =\int_0^\infty e^{-i\omega t}\frac{i}{2N}\langle\Omega\vert T\rho_{\mathbf{q}}(t)\rho_{-\mathbf{q}}(0)\vert\Omega\rangle dt,\\
\chi^{ij}_0(\mathbf{q},\omega)  & = \int_0^\infty   e^{-i\omega t}\frac{i}{2N}\langle\Omega\vert TS^i_{\mathbf{q}}(t)S^j_{-\mathbf{q}}(0)\vert\Omega\rangle dt.
\end{align}
In details, we can conduct the derivations
\begin{align}
\chi_{0}^{00}(\mathbf{q},\omega) & =\int_0^\infty dt e^{-i\omega t}\frac{i}{2N}\langle\Omega\vert T\rho_{\mathbf{q}}(t)\rho_{-\mathbf{q}}(0)\vert\Omega\rangle \notag \\
 & =-\frac{1}{2N}\sum_{\mathbf{k}}\sum_{\sigma} \frac{n_{\mathbf{k}+\mathbf{q},\sigma} -n_{\mathbf{k},\sigma}}{\omega+\xi_{\mathbf{k},\sigma}-\xi_{\mathbf{k}+\mathbf{q},\sigma}},
\end{align}
and 
\begin{equation}
{\small
\begin{split}
&\chi_{0}^{ij}(\mathbf{q},\omega) \\
=&\sum_{\mathbf k\mathbf k^\prime}\sum_{\alpha\beta}\sum_{\alpha^{\prime}\beta^{\prime}}\langle\Omega | T c_{\mathbf{k}+\mathbf q,\alpha}^{\dagger}c_{\mathbf{k},\beta}\sigma_{\alpha\beta}^{i}c_{\mathbf{k}^{\prime}-\mathbf q,\alpha^{\prime}}^{\dagger}c_{\mathbf{k}^{\prime},\beta^{\prime}}\sigma_{\alpha^{\prime}\beta^{\prime}}^{j}|\Omega \rangle \notag \\
=&-\sum_{\mathbf k\mathbf k^\prime}\sum_{\alpha\beta}\sum_{\alpha^{\prime}\beta^{\prime}}\sigma_{\alpha\beta}^{i}\sigma_{\alpha^{\prime}\beta^{\prime}}^{j}\langle\Omega | c_{\mathbf{k}+\mathbf q,\alpha}^{\dagger}c_{\mathbf{k}^{\prime},\beta^{\prime}}|\Omega\rangle\langle\Omega|c_{\mathbf{k}^{\prime}-\mathbf q,\alpha^{\prime}}^{\dagger} c_{\mathbf{k},\beta}|\Omega\rangle \notag \\
=&-\sum_{\mathbf k}\sum_{\alpha\beta}\sum_{\alpha^{\prime}\beta^{\prime}} \sigma_{\alpha\beta}^{i}\sigma_{\alpha^{\prime}\beta^{\prime}}^{j}\delta_{\alpha\beta^{\prime}}\delta_{\alpha^{\prime}\beta}\frac{n_\mathrm{FD}(\xi_{\mathbf{k}+\mathbf{q},\alpha})-n_\mathrm{FD}(\xi_{\mathbf{k},\beta})}{\omega+\xi_{\mathbf{k},\alpha}-\xi_{\mathbf{k}+\mathbf{q},\beta}} \notag \\
=&-\delta^{ij}\sum_{\mathbf k}\sum_{\sigma}\frac{n_\mathrm{FD}(\xi_{\mathbf{k},\sigma})-n_\mathrm{FD}(\xi_{\mathbf{k}+\mathbf{q},\bar{\sigma}})}{\omega+\xi_{\mathbf{k},\sigma}-\xi_{\mathbf{k}+\mathbf{q},\bar{\sigma}}}
 \notag  \\
=& \begin{cases}
\chi_{0}^{00}(\mathbf{q},\omega) & i=j=z\\
\delta_{ij}\chi_{0}^{+-}(\mathbf{q},\omega) & i,j=x,y
\end{cases}
\end{split}
}
\end{equation}
where $n_\mathrm{FD}(\xi)= 1/(e^{\beta \xi}+1)$ is the Fermi-Dirac distribution. To further obtain the spin wave, we can utilize RPA, which leads the susceptibility to be
\begin{align}
\chi^{00}_\mathrm{RPA}(\mathbf{q},\omega) & =\frac{\chi_{0}^{00}(\mathbf{q},\omega)}{1-U\chi_{0}^{00}(\mathbf{q},\omega)}\label{eqS:RPA_rho},\\
\chi^{xx,yy}_\mathrm{RPA}(\mathbf{q},\omega) & =\frac{\chi_{0}^{+-}(\mathbf{q},\omega)}{1 - U\chi_{0}^{+-}(\mathbf{q},\omega)}\label{eqS:RPA_spin}~.
\end{align}
with the denominators
\begin{align}
&1 - U\chi_{0}^{00}(\mathbf{q},\omega) \notag\\ =&1+\frac{U}{2N}\sum_{\mathbf{k}}\sum_{\sigma}\frac{n_\mathrm{FD}(\xi_{\mathbf{k},\sigma}) -n_\mathrm{FD}(\xi_{\mathbf{k}+\mathbf{q},\sigma})}{\omega+\xi_{\mathbf{k},\sigma} -\xi_{\mathbf{k}+\mathbf{q},\sigma}} \\
&1 - U\chi_{0}^{+-}(\mathbf{q},\omega) \notag\\ =&1+\frac{U}{2N}\sum_{\mathbf{k}}\sum_{\sigma}\frac{n_\mathrm{FD}(\xi_{\mathbf{k},\sigma}) - n_\mathrm{FD}(\xi_{\mathbf{k}+\mathbf{q},\bar{\sigma}})}{\omega+\xi_{\mathbf{k},\sigma} -\xi_{\mathbf{k}+\mathbf{q},\bar{\sigma}}}.
\end{align}
In particular, at $q=0$, the conditions where the two denominators vanish recover the self-consistent equation for the Hartree-Fock parameters,
\begin{align}
\frac{1}{U} & =-\frac{1}{2} \sum_{\mathbf{k}}\sum_{\sigma}\lim_{q\rightarrow0}\frac{n_\mathrm{FD}(\xi_{\mathbf{k},\sigma}) -n_\mathrm{FD}(\xi_{\mathbf{k}+\mathbf{q},\sigma})}{\xi_{\mathbf{k},\sigma} -\xi_{\mathbf{k}+\mathbf{q},\sigma}}\notag\\ &=-\frac{1}{N}\sum_{\mathbf{k}}\sum_{\sigma}\partial_{\epsilon}n_\mathrm{FD}(\xi_{\mathbf{k},\sigma}), \\
\frac{1}{U} & =- \frac{1}{2}\sum_{\text{\ensuremath{\mathbf{k}}}}\sum_{\sigma} \lim_{q\rightarrow0}\frac{n_\mathrm{FD}(\xi_{\mathbf{k},\sigma}) -n_\mathrm{FD}(\xi_{\mathbf{k}+\mathbf{q},\bar{\sigma}})}{\xi_{\mathbf{k},\sigma} -\xi_{\mathbf{k}+\mathbf{q},\bar{\sigma}}}\notag\\ &=\frac{1}{N}\sum_{\mathbf{k}}\frac{n_\mathrm{FD}(\xi_{\mathbf{k},\uparrow}) -n_\mathrm{FD}(\xi_{\mathbf{k},\downarrow})}{Um}.
\end{align}
Here $\partial_\epsilon n_\mathrm{FD}(\xi) \equiv \frac{\partial}{\partial \xi} n_\mathrm{FD}(\xi)$. 
The two equations determine the chemical potential and the magnetization, respectively.

The spin wave dispersion is determined from the pole of the spin susceptibility $\chi_\mathrm{RPA}^{ij}(\mathbf{q},\omega)$.
We note that
\begin{align}
\chi_{0}^{+-}(\mathbf{q},\omega)= & -\frac{1}{2N}\sum_{\mathbf{k}}\sum_{\sigma}\frac{n_\mathrm{FD}(\xi_{\mathbf{k},\sigma}) -n_\mathrm{FD}(\xi_{\mathbf{k}+\mathbf{q},\bar{\sigma}})}{\omega+\xi_{\mathbf{k},\sigma} -\xi_{\mathbf{k}+\mathbf{q},\bar{\sigma}}}\nonumber \\
= &  -\frac{1}{2N}\sum_{\mathbf k}\sum_{\sigma}\frac{n_\mathrm{FD}(\xi_{\mathbf{k},\sigma}) -n_\mathrm{FD}(\xi_{\mathbf{k}+\mathbf{q},\bar{\sigma}})}{\omega+\epsilon_{\mathbf{k}} -\epsilon_{\mathbf{k}+\mathbf{q}}-Um}\nonumber \\
\simeq &  \frac{1}{U}+\frac{\omega}{mU^{2}}-\frac{q^{2}}{2m_{e}mU^{2}},
\end{align}
where we make the approximation $\epsilon_{\mathbf{k}}=\frac{\mathbf{k}^{2}}{2m_{e}}$
with $m_{e}$ being the effective mass of the electron. We obtain the spin
susceptibility,
\begin{equation}
\chi_\mathrm{RPA}^{ij}(\mathbf{q},\omega) = - \frac{\chi_{0}^{+-}(\mathbf{q},\omega)}{\frac{\omega}{mU}-\frac{q^{2}}{2m_{e}mU}}\delta_{ij}.
\end{equation}
Therefore, the spin wave possesses the square dispersion $\omega(\mathbf q) \propto q^2$.

\subsection{Spin wave gap from magnetic anisotropy}
Indeed, the introduction of magnetic anisotropy, which breaks the $SU(2)$ symmetry, can help stabilize the FM. 
A simple example of this is obtained by adding the anisotropic term \eqref{eqS:J3} to the Hubbard Hamiltonian.
The $\delta H$ commutes with the Hubbard model Hamiltonian. Thus, the extra contribution to the Hartree-Fock Hamiltonian 
\begin{equation}
\delta H_\mathrm{HF} = - 2zmJ_{3}\sum_{\mathbf{i}}S_{\mathbf{i}}^{z} + zm^{2}J_{3}
\end{equation}
does not alter the Hartree-Fock ground state. Here, $z$ is the coordinate number.  
The spin susceptibility accordingly
turns out to be
\begin{align}
\chi_{0}^{ij}(\mathbf{q},\omega) &  \simeq \frac{1}{U}+\frac{\omega}{mU^{2}}-\frac{q^{2}/(2m_e)+2zJ_{3}}{mU^{2}},
\end{align}
and 
\begin{equation}
\chi_\mathrm{RPA}^{ij}(\mathbf{q},\omega)=-\frac{\chi_{0}^{+-}(\mathbf{q},\omega)}{\frac{\omega}{mU}-\frac{q^{2}/(2m_e)+2zJ_{3}}{mU} }\delta_{ij}.
\end{equation}
which indicates a finite gap in Eq.~\eqref{eqS:gap} of the spin wave.  
This gap arises from the breaking of $SU(2)$ symmetry due to the magnetic anisotropy. 
As a consequence, the finite gap $\Delta$ can suppress the thermal fluctuations of spin wave as such to stabilize a FM at finite temperature.
The above RPA analysis can be applied to a general magnetic anisotropic term which can induce a spin-wave gap due to the breakdown of $SU(2)$ symmetry.



\begin{thebibliography}{10}
\expandafter\ifx\csname url\endcsname\relax
  \def\url#1{\texttt{#1}}\fi
\expandafter\ifx\csname urlprefix\endcsname\relax\def\urlprefix{URL }\fi
\providecommand{\bibinfo}[2]{#2}
\providecommand{\eprint}[2][]{\url{#2}}

\bibitem{Gutzwiller1963}
\bibinfo{author}{Gutzwiller, M.~C.}
\newblock \bibinfo{title}{{Effect of Correlation on the Ferromagnetism of
  Transition Metals}}.
\newblock \emph{\bibinfo{journal}{Phys. Rev. Lett.}}
  \textbf{\bibinfo{volume}{10}}, \bibinfo{pages}{159--162}
  (\bibinfo{year}{1963}).
\newblock \urlprefix\url{https://link.aps.org/doi/10.1103/PhysRevLett.10.159}.

\bibitem{Hubbard1963}
\bibinfo{author}{Hubbard, J.} \& \bibinfo{author}{Flowers, B.~H.}
\newblock \bibinfo{title}{{Electron correlations in narrow energy bands}}.
\newblock \emph{\bibinfo{journal}{Proceedings of the Royal Society of London.
  Series A. Mathematical and Physical Sciences}}
  \textbf{\bibinfo{volume}{276}}, \bibinfo{pages}{238--257}
  (\bibinfo{year}{1963}).
\newblock
  \urlprefix\url{https://royalsocietypublishing.org/doi/abs/10.1098/rspa.1963.0204}.

\bibitem{Tasaki1998}
\bibinfo{author}{Tasaki, H.}
\newblock \bibinfo{title}{{The Hubbard model - an introduction and selected
  rigorous results}}.
\newblock \emph{\bibinfo{journal}{Journal of Physics: Condensed Matter}}
  \textbf{\bibinfo{volume}{10}}, \bibinfo{pages}{4353} (\bibinfo{year}{1998}).
\newblock \urlprefix\url{https://dx.doi.org/10.1088/0953-8984/10/20/004}.

\bibitem{Tasaki1998Review}
\bibinfo{author}{Tasaki, H.}
\newblock \bibinfo{title}{{From Nagaoka's Ferromagnetism to Flat-Band
  Ferromagnetism and Beyond: An Introduction to Ferromagnetism in the Hubbard
  Model}}.
\newblock \emph{\bibinfo{journal}{Progress of Theoretical Physics}}
  \textbf{\bibinfo{volume}{99}}, \bibinfo{pages}{489--548}
  (\bibinfo{year}{1998}).
\newblock \urlprefix\url{https://doi.org/10.1143/PTP.99.489}.

\bibitem{Vollhardt1999}
\bibinfo{author}{Vollhardt, D.} \emph{et~al.}
\newblock \bibinfo{title}{Metallic ferromagnetism: Progress in our
  understanding of an old strong-coupling problem}.
\newblock In \bibinfo{editor}{Kramer, B.} (ed.)
  \emph{\bibinfo{booktitle}{Advances in Solid State Physics 38}},
  \bibinfo{pages}{383--396} (\bibinfo{publisher}{Springer Berlin Heidelberg},
  \bibinfo{address}{Berlin, Heidelberg}, \bibinfo{year}{1999}).

\bibitem{Stoner1938}
\bibinfo{author}{Stoner, E.~C.}
\newblock \bibinfo{title}{{Collective electron ferromagnetism}}.
\newblock \emph{\bibinfo{journal}{Proceedings of the Royal Society of London.
  Series A. Mathematical and Physical Sciences}}
  \textbf{\bibinfo{volume}{165}}, \bibinfo{pages}{372--414}
  (\bibinfo{year}{1938}).
\newblock
  \urlprefix\url{https://royalsocietypublishing.org/doi/abs/10.1098/rspa.1938.0066}.

\bibitem{Vollhardt1996}
\bibinfo{author}{Vollhardt, D.} \emph{et~al.}
\newblock \bibinfo{title}{{Non-perturbative approaches to magnetism in strongly
  correlated electron systems}}.
\newblock \emph{\bibinfo{journal}{Zeitschrift f{\"u}r Physik B Condensed
  Matter}} \textbf{\bibinfo{volume}{103}}, \bibinfo{pages}{283--292}
  (\bibinfo{year}{1996}).

\bibitem{Thouless1965}
\bibinfo{author}{Thouless, D.~J.}
\newblock \bibinfo{title}{{Exchange in solid 3He and the Heisenberg
  Hamiltonian}}.
\newblock \emph{\bibinfo{journal}{Proceedings of the Physical Society}}
  \textbf{\bibinfo{volume}{86}}, \bibinfo{pages}{893} (\bibinfo{year}{1965}).
\newblock \urlprefix\url{https://dx.doi.org/10.1088/0370-1328/86/5/301}.

\bibitem{Nagaoka1966}
\bibinfo{author}{Nagaoka, Y.}
\newblock \bibinfo{title}{{Ferromagnetism in a Narrow, Almost Half-Filled $s$
  Band}}.
\newblock \emph{\bibinfo{journal}{Phys. Rev.}} \textbf{\bibinfo{volume}{147}},
  \bibinfo{pages}{392--405} (\bibinfo{year}{1966}).
\newblock \urlprefix\url{https://link.aps.org/doi/10.1103/PhysRev.147.392}.

\bibitem{Mielke1991b}
\bibinfo{author}{Mielke, A.}
\newblock \bibinfo{title}{{Ferromagnetism in the Hubbard model on line graphs
  and further considerations}}.
\newblock \emph{\bibinfo{journal}{Journal of Physics A: Mathematical and
  General}} \textbf{\bibinfo{volume}{24}}, \bibinfo{pages}{3311}
  (\bibinfo{year}{1991}).
\newblock \urlprefix\url{https://dx.doi.org/10.1088/0305-4470/24/14/018}.

\bibitem{Tasaki1992}
\bibinfo{author}{Tasaki, H.}
\newblock \bibinfo{title}{Ferromagnetism in the hubbard models with degenerate
  single-electron ground states}.
\newblock \emph{\bibinfo{journal}{Phys. Rev. Lett.}}
  \textbf{\bibinfo{volume}{69}}, \bibinfo{pages}{1608--1611}
  (\bibinfo{year}{1992}).
\newblock \urlprefix\url{https://link.aps.org/doi/10.1103/PhysRevLett.69.1608}.

\bibitem{Riera1989}
\bibinfo{author}{Riera, J.~A.} \& \bibinfo{author}{Young, A.~P.}
\newblock \bibinfo{title}{{Ferromagnetism in the one-band Hubbard model}}.
\newblock \emph{\bibinfo{journal}{Phys. Rev. B}} \textbf{\bibinfo{volume}{40}},
  \bibinfo{pages}{5285--5288} (\bibinfo{year}{1989}).
\newblock \urlprefix\url{https://link.aps.org/doi/10.1103/PhysRevB.40.5285}.

\bibitem{Basile1990}
\bibinfo{author}{Basile, A.~G.} \& \bibinfo{author}{Elser, V.}
\newblock \bibinfo{title}{{Stability of the ferromagnetic state with respect to
  a single spin flip: Variational calculations for the U=\ensuremath{\infty}
  Hubbard model on the square lattice}}.
\newblock \emph{\bibinfo{journal}{Phys. Rev. B}} \textbf{\bibinfo{volume}{41}},
  \bibinfo{pages}{4842--4845} (\bibinfo{year}{1990}).
\newblock \urlprefix\url{https://link.aps.org/doi/10.1103/PhysRevB.41.4842}.

\bibitem{Emery1990}
\bibinfo{author}{Emery, V.~J.}, \bibinfo{author}{Kivelson, S.~A.} \&
  \bibinfo{author}{Lin, H.~Q.}
\newblock \bibinfo{title}{{Phase separation in the t-J model}}.
\newblock \emph{\bibinfo{journal}{Phys. Rev. Lett.}}
  \textbf{\bibinfo{volume}{64}}, \bibinfo{pages}{475--478}
  (\bibinfo{year}{1990}).
\newblock \urlprefix\url{https://link.aps.org/doi/10.1103/PhysRevLett.64.475}.

\bibitem{Shastry1990}
\bibinfo{author}{Shastry, B.~S.}, \bibinfo{author}{Krishnamurthy, H.~R.} \&
  \bibinfo{author}{Anderson, P.~W.}
\newblock \bibinfo{title}{Instability of the nagaoka ferromagnetic state of the
  u=\ensuremath{\infty} hubbard model}.
\newblock \emph{\bibinfo{journal}{Phys. Rev. B}} \textbf{\bibinfo{volume}{41}},
  \bibinfo{pages}{2375--2379} (\bibinfo{year}{1990}).
\newblock \urlprefix\url{https://link.aps.org/doi/10.1103/PhysRevB.41.2375}.

\bibitem{Mielke1991}
\bibinfo{author}{Mielke, A.}
\newblock \bibinfo{title}{{Ferromagnetic ground states for the Hubbard model on
  line graphs}}.
\newblock \emph{\bibinfo{journal}{Journal of Physics A: Mathematical and
  General}} \textbf{\bibinfo{volume}{24}}, \bibinfo{pages}{L73}
  (\bibinfo{year}{1991}).
\newblock \urlprefix\url{https://dx.doi.org/10.1088/0305-4470/24/2/005}.

\bibitem{Mielke1992}
\bibinfo{author}{Mielke, A.}
\newblock \bibinfo{title}{{Exact ground states for the Hubbard model on the
  Kagome lattice}}.
\newblock \emph{\bibinfo{journal}{Journal of Physics A: Mathematical and
  General}} \textbf{\bibinfo{volume}{25}}, \bibinfo{pages}{4335}
  (\bibinfo{year}{1992}).
\newblock \urlprefix\url{https://dx.doi.org/10.1088/0305-4470/25/16/011}.

\bibitem{Hanisch1995}
\bibinfo{author}{Hanisch, T.}, \bibinfo{author}{Kleine, B.},
  \bibinfo{author}{Ritzl, A.} \& \bibinfo{author}{Müller-Hartmann, E.}
\newblock \bibinfo{title}{{Ferromagnetism in the Hubbard model: instability of
  the Nagaoka state on the triangular, honeycomb and kagome lattices}}.
\newblock \emph{\bibinfo{journal}{Annalen der Physik}}
  \textbf{\bibinfo{volume}{507}}, \bibinfo{pages}{303--328}
  (\bibinfo{year}{1995}).
\newblock
  \urlprefix\url{https://onlinelibrary.wiley.com/doi/abs/10.1002/andp.19955070405}.

\bibitem{Wurth1996}
\bibinfo{author}{Wurth, P.}, \bibinfo{author}{Uhrig, G.} \&
  \bibinfo{author}{Müller-Hartmann, E.}
\newblock \bibinfo{title}{{Ferromagnetism in the Hubbard model on the square
  lattice: Improved instability criterion for the Nagaoka state}}.
\newblock \emph{\bibinfo{journal}{Annalen der Physik}}
  \textbf{\bibinfo{volume}{508}}, \bibinfo{pages}{148--155}
  (\bibinfo{year}{1996}).
\newblock
  \urlprefix\url{https://onlinelibrary.wiley.com/doi/abs/10.1002/andp.2065080204}.

\bibitem{Brunner1998}
\bibinfo{author}{Brunner, M.} \& \bibinfo{author}{Muramatsu, A.}
\newblock \bibinfo{title}{Quantum monte carlo simulations of infinitely
  strongly correlated fermions}.
\newblock \emph{\bibinfo{journal}{Phys. Rev. B}} \textbf{\bibinfo{volume}{58}},
  \bibinfo{pages}{R10100--R10103} (\bibinfo{year}{1998}).
\newblock \urlprefix\url{https://link.aps.org/doi/10.1103/PhysRevB.58.R10100}.

\bibitem{Ulmke1998}
\bibinfo{author}{Ulmke, M.}
\newblock \bibinfo{title}{Ferromagnetism in the hubbard model on fcc-type
  lattices}.
\newblock \emph{\bibinfo{journal}{The European Physical Journal B-Condensed
  Matter and Complex Systems}} \textbf{\bibinfo{volume}{1}},
  \bibinfo{pages}{301--304} (\bibinfo{year}{1998}).

\bibitem{Becca2001}
\bibinfo{author}{Becca, F.} \& \bibinfo{author}{Sorella, S.}
\newblock \bibinfo{title}{Nagaoka ferromagnetism in the two-dimensional
  infinite- $\mathit{U}$ hubbard model}.
\newblock \emph{\bibinfo{journal}{Phys. Rev. Lett.}}
  \textbf{\bibinfo{volume}{86}}, \bibinfo{pages}{3396--3399}
  (\bibinfo{year}{2001}).
\newblock \urlprefix\url{https://link.aps.org/doi/10.1103/PhysRevLett.86.3396}.

\bibitem{Merino2006}
\bibinfo{author}{Merino, J.}, \bibinfo{author}{Powell, B.~J.} \&
  \bibinfo{author}{McKenzie, R.~H.}
\newblock \bibinfo{title}{{Ferromagnetism, paramagnetism, and a Curie-Weiss
  metal in an electron-doped Hubbard model on a triangular lattice}}.
\newblock \emph{\bibinfo{journal}{Phys. Rev. B}} \textbf{\bibinfo{volume}{73}},
  \bibinfo{pages}{235107} (\bibinfo{year}{2006}).
\newblock \urlprefix\url{https://link.aps.org/doi/10.1103/PhysRevB.73.235107}.

\bibitem{ChiaChenChang2010}
\bibinfo{author}{Chang, C.-C.}, \bibinfo{author}{Zhang, S.} \&
  \bibinfo{author}{Ceperley, D.~M.}
\newblock \bibinfo{title}{{Itinerant ferromagnetism in a Fermi gas with contact
  interaction: Magnetic properties in a dilute Hubbard model}}.
\newblock \emph{\bibinfo{journal}{Phys. Rev. A}} \textbf{\bibinfo{volume}{82}},
  \bibinfo{pages}{061603} (\bibinfo{year}{2010}).
\newblock \urlprefix\url{https://link.aps.org/doi/10.1103/PhysRevA.82.061603}.

\bibitem{SoonYongChang2011}
\bibinfo{author}{Chang, S.-Y.}, \bibinfo{author}{Randeria, M.} \&
  \bibinfo{author}{Trivedi, N.}
\newblock \bibinfo{title}{{Ferromagnetism in the upper branch of the Feshbach
  resonance and the hard-sphere Fermi gas}}.
\newblock \emph{\bibinfo{journal}{Proceedings of the National Academy of
  Sciences}} \textbf{\bibinfo{volume}{108}}, \bibinfo{pages}{51--54}
  (\bibinfo{year}{2011}).
\newblock \urlprefix\url{https://www.pnas.org/doi/abs/10.1073/pnas.1011990108}.

\bibitem{Carleo2011}
\bibinfo{author}{Carleo, G.}, \bibinfo{author}{Moroni, S.},
  \bibinfo{author}{Becca, F.} \& \bibinfo{author}{Baroni, S.}
\newblock \bibinfo{title}{Itinerant ferromagnetic phase of the hubbard model}.
\newblock \emph{\bibinfo{journal}{Phys. Rev. B}} \textbf{\bibinfo{volume}{83}},
  \bibinfo{pages}{060411} (\bibinfo{year}{2011}).
\newblock \urlprefix\url{https://link.aps.org/doi/10.1103/PhysRevB.83.060411}.

\bibitem{LiLiu2012}
\bibinfo{author}{Liu, L.}, \bibinfo{author}{Yao, H.}, \bibinfo{author}{Berg,
  E.}, \bibinfo{author}{White, S.~R.} \& \bibinfo{author}{Kivelson, S.~A.}
\newblock \bibinfo{title}{{Phases of the Infinite $U$ Hubbard Model on Square
  Lattices}}.
\newblock \emph{\bibinfo{journal}{Phys. Rev. Lett.}}
  \textbf{\bibinfo{volume}{108}}, \bibinfo{pages}{126406}
  (\bibinfo{year}{2012}).
\newblock
  \urlprefix\url{https://link.aps.org/doi/10.1103/PhysRevLett.108.126406}.

\bibitem{GangLi2014}
\bibinfo{author}{Li, G.}, \bibinfo{author}{Antipov, A.~E.},
  \bibinfo{author}{Rubtsov, A.~N.}, \bibinfo{author}{Kirchner, S.} \&
  \bibinfo{author}{Hanke, W.}
\newblock \bibinfo{title}{{Competing phases of the Hubbard model on a
  triangular lattice: Insights from the entropy}}.
\newblock \emph{\bibinfo{journal}{Phys. Rev. B}} \textbf{\bibinfo{volume}{89}},
  \bibinfo{pages}{161118} (\bibinfo{year}{2014}).
\newblock \urlprefix\url{https://link.aps.org/doi/10.1103/PhysRevB.89.161118}.

\bibitem{Schlomer2023}
\bibinfo{author}{Schlömer, H.}, \bibinfo{author}{Schollwöck, U.},
  \bibinfo{author}{Bohrdt, A.} \& \bibinfo{author}{Grusdt, F.}
\newblock \bibinfo{title}{{Kinetic-to-magnetic frustration crossover and linear
  confinement in the doped triangular $t-J$ model}} (\bibinfo{year}{2023}).
\newblock \eprint{2305.02342}.

\bibitem{Davydova2023}
\bibinfo{author}{Davydova, M.}, \bibinfo{author}{Zhang, Y.} \&
  \bibinfo{author}{Fu, L.}
\newblock \bibinfo{title}{Itinerant spin polaron and metallic ferromagnetism in
  semiconductor moir\'e superlattices}.
\newblock \emph{\bibinfo{journal}{Phys. Rev. B}}
  \textbf{\bibinfo{volume}{107}}, \bibinfo{pages}{224420}
  (\bibinfo{year}{2023}).
\newblock \urlprefix\url{https://link.aps.org/doi/10.1103/PhysRevB.107.224420}.

\bibitem{Samajdar2023}
\bibinfo{author}{Samajdar, R.} \& \bibinfo{author}{Bhatt, R.~N.}
\newblock \bibinfo{title}{Nagaoka ferromagnetism in doped hubbard models in
  optical lattices} (\bibinfo{year}{2023}).
\newblock \eprint{2305.05683}.

\bibitem{YuchiHe2023}
\bibinfo{author}{He, Y.} \emph{et~al.}
\newblock \bibinfo{title}{{Itinerant Magnetism in the Triangular Lattice
  Hubbard Model at Half-doping: Application to Twisted Transition-Metal
  Dichalcogenides}} (\bibinfo{year}{2023}).
\newblock \eprint{2311.10146}.

\bibitem{Morera2024}
\bibinfo{author}{Morera, I.} \& \bibinfo{author}{Demler, E.}
\newblock \bibinfo{title}{{Itinerant magnetism and magnetic polarons in the
  triangular lattice Hubbard model}} (\bibinfo{year}{2024}).
\newblock \eprint{2402.14074}.

\bibitem{Lorenzo2024}
\bibinfo{author}{Del~Re, L.} \& \bibinfo{author}{Classen, L.}
\newblock \bibinfo{title}{Field control of symmetry-broken and quantum
  disordered phases in frustrated moir\'e bilayers with population imbalance}.
\newblock \emph{\bibinfo{journal}{Phys. Rev. Res.}}
  \textbf{\bibinfo{volume}{6}}, \bibinfo{pages}{023082} (\bibinfo{year}{2024}).
\newblock
  \urlprefix\url{https://link.aps.org/doi/10.1103/PhysRevResearch.6.023082}.

\bibitem{Samajdar2024PRB}
\bibinfo{author}{Samajdar, R.} \& \bibinfo{author}{Bhatt, R.~N.}
\newblock \bibinfo{title}{Polaronic mechanism of nagaoka ferromagnetism in
  hubbard models}.
\newblock \emph{\bibinfo{journal}{Phys. Rev. B}}
  \textbf{\bibinfo{volume}{109}}, \bibinfo{pages}{235128}
  (\bibinfo{year}{2024}).
\newblock \urlprefix\url{https://link.aps.org/doi/10.1103/PhysRevB.109.235128}.

\bibitem{YanhaoTang2020}
\bibinfo{author}{Tang, Y.} \emph{et~al.}
\newblock \bibinfo{title}{{Simulation of Hubbard model physics in WSe2/WS2
  moir{\'e} superlattices}}.
\newblock \emph{\bibinfo{journal}{Nature}} \textbf{\bibinfo{volume}{579}},
  \bibinfo{pages}{353--358} (\bibinfo{year}{2020}).
\newblock \urlprefix\url{https://www.nature.com/articles/s41586-020-2085-3}.

\bibitem{Ciorciaro2023}
\bibinfo{author}{Ciorciaro, L.} \emph{et~al.}
\newblock \bibinfo{title}{{Kinetic magnetism in triangular moir{\'e}
  materials}}.
\newblock \emph{\bibinfo{journal}{Nature}} \textbf{\bibinfo{volume}{623}},
  \bibinfo{pages}{509--513} (\bibinfo{year}{2023}).
\newblock \urlprefix\url{https://www.nature.com/articles/s41586-023-06633-0}.

\bibitem{Anderson2023}
\bibinfo{author}{Anderson, E.} \emph{et~al.}
\newblock \bibinfo{title}{{Programming correlated magnetic states with
  gate-controlled moiré geometry}}.
\newblock \emph{\bibinfo{journal}{Science}} \textbf{\bibinfo{volume}{381}},
  \bibinfo{pages}{325--330} (\bibinfo{year}{2023}).
\newblock
  \urlprefix\url{https://www.science.org/doi/abs/10.1126/science.adg4268}.

\bibitem{Seifert2024}
\bibinfo{author}{Seifert, U. F.~P.} \& \bibinfo{author}{Balents, L.}
\newblock \bibinfo{title}{Spin polarons and ferromagnetism in doped dilute
  moir\'e-mott insulators}.
\newblock \emph{\bibinfo{journal}{Phys. Rev. Lett.}}
  \textbf{\bibinfo{volume}{132}}, \bibinfo{pages}{046501}
  (\bibinfo{year}{2024}).
\newblock
  \urlprefix\url{https://link.aps.org/doi/10.1103/PhysRevLett.132.046501}.

\bibitem{YangJin2021}
\bibinfo{author}{Yang, J.}, \bibinfo{author}{Liu, L.},
  \bibinfo{author}{Mongkolkiattichai, J.} \& \bibinfo{author}{Schauss, P.}
\newblock \bibinfo{title}{Site-resolved imaging of ultracold fermions in a
  triangular-lattice quantum gas microscope}.
\newblock \emph{\bibinfo{journal}{PRX Quantum}} \textbf{\bibinfo{volume}{2}},
  \bibinfo{pages}{020344} (\bibinfo{year}{2021}).
\newblock \urlprefix\url{https://link.aps.org/doi/10.1103/PRXQuantum.2.020344}.

\bibitem{MuqingXu2023}
\bibinfo{author}{Xu, M.} \emph{et~al.}
\newblock \bibinfo{title}{{Frustration-and doping-induced magnetism in a
  fermi--hubbard simulator}}.
\newblock \emph{\bibinfo{journal}{Nature}} \textbf{\bibinfo{volume}{620}},
  \bibinfo{pages}{971--976} (\bibinfo{year}{2023}).
\newblock \urlprefix\url{https://www.nature.com/articles/s41586-023-06280-5}.

\bibitem{Lebrat2024}
\bibinfo{author}{Lebrat, M.} \emph{et~al.}
\newblock \bibinfo{title}{Observation of nagaoka polarons in a fermi--hubbard
  quantum simulator}.
\newblock \emph{\bibinfo{journal}{Nature}} \textbf{\bibinfo{volume}{629}},
  \bibinfo{pages}{317--322} (\bibinfo{year}{2024}).
\newblock \urlprefix\url{https://www.nature.com/articles/s41586-024-07272-9}.

\bibitem{Prichard2023}
\bibinfo{author}{Prichard, M.~L.} \emph{et~al.}
\newblock \bibinfo{title}{Directly imaging spin polarons in a kinetically
  frustrated hubbard system}.
\newblock \emph{\bibinfo{journal}{Nature}} \textbf{\bibinfo{volume}{629}},
  \bibinfo{pages}{323--328} (\bibinfo{year}{2024}).
\newblock \urlprefix\url{https://www.nature.com/articles/s41586-024-07356-6}.

\bibitem{ChinCheng2010}
\bibinfo{author}{Chin, C.}, \bibinfo{author}{Grimm, R.},
  \bibinfo{author}{Julienne, P.} \& \bibinfo{author}{Tiesinga, E.}
\newblock \bibinfo{title}{Feshbach resonances in ultracold gases}.
\newblock \emph{\bibinfo{journal}{Rev. Mod. Phys.}}
  \textbf{\bibinfo{volume}{82}}, \bibinfo{pages}{1225--1286}
  (\bibinfo{year}{2010}).
\newblock \urlprefix\url{https://link.aps.org/doi/10.1103/RevModPhys.82.1225}.

\bibitem{Lewenstein2012}
\bibinfo{author}{Lewenstein, M.}, \bibinfo{author}{Sanpera, A.} \&
  \bibinfo{author}{Ahufinger, V.}
\newblock \emph{\bibinfo{title}{Ultracold Atoms in Optical Lattices: Simulating
  quantum many-body systems}} (\bibinfo{publisher}{Oxford University Press},
  \bibinfo{year}{2012}).

\bibitem{Zhu2023SC}
\bibinfo{author}{Zhu, Z.} \& \bibinfo{author}{Chen, Q.}
\newblock \bibinfo{title}{Superconductivity in doped triangular mott
  insulators: The roles of parent spin backgrounds and charge kinetic energy}.
\newblock \emph{\bibinfo{journal}{Phys. Rev. B}}
  \textbf{\bibinfo{volume}{107}}, \bibinfo{pages}{L220502}
  (\bibinfo{year}{2023}).
\newblock
  \urlprefix\url{https://link.aps.org/doi/10.1103/PhysRevB.107.L220502}.

\bibitem{ShuaiChen2022SCBA}
\bibinfo{author}{Chen, S.~A.}, \bibinfo{author}{Chen, Q.} \&
  \bibinfo{author}{Zhu, Z.}
\newblock \bibinfo{title}{Proposal for asymmetric photoemission and tunneling
  spectroscopies in quantum simulators of the triangular-lattice
  {Fermi}-{Hubbard} model}.
\newblock \emph{\bibinfo{journal}{Phys. Rev. B}}
  \textbf{\bibinfo{volume}{106}}, \bibinfo{pages}{085138}
  (\bibinfo{year}{2022}).

\bibitem{GuanhuaHuang2023}
\bibinfo{author}{hua Huang, G.} \& \bibinfo{author}{Wu, Z.}
\newblock \bibinfo{title}{{Magnetic correlations of a doped and frustrated
  Hubbard model: benchmarking the two-particle self-consistent theory against a
  quantum simulator}} (\bibinfo{year}{2023}).
\newblock \eprint{2310.11263}.

\bibitem{Morera2023}
\bibinfo{author}{Morera, I.} \emph{et~al.}
\newblock \bibinfo{title}{High-temperature kinetic magnetism in triangular
  lattices}.
\newblock \emph{\bibinfo{journal}{Phys. Rev. Res.}}
  \textbf{\bibinfo{volume}{5}}, \bibinfo{pages}{L022048}
  (\bibinfo{year}{2023}).
\newblock
  \urlprefix\url{https://link.aps.org/doi/10.1103/PhysRevResearch.5.L022048}.

\bibitem{KyungminLee2023}
\bibinfo{author}{Lee, K.}, \bibinfo{author}{Sharma, P.},
  \bibinfo{author}{Vafek, O.} \& \bibinfo{author}{Changlani, H.~J.}
\newblock \bibinfo{title}{Triangular lattice hubbard model physics at
  intermediate temperatures}.
\newblock \emph{\bibinfo{journal}{Phys. Rev. B}}
  \textbf{\bibinfo{volume}{107}}, \bibinfo{pages}{235105}
  (\bibinfo{year}{2023}).
\newblock \urlprefix\url{https://link.aps.org/doi/10.1103/PhysRevB.107.235105}.

\bibitem{ChengshuLi2023}
\bibinfo{author}{Li, C.}, \bibinfo{author}{He, M.-G.}, \bibinfo{author}{Wang,
  C.-Y.} \& \bibinfo{author}{Zhai, H.}
\newblock \bibinfo{title}{{Frustration induced Itinerant Ferromagnetism of
  Fermions in Optical Lattice}} (\bibinfo{year}{2023}).
\newblock \eprint{2305.01682}.

\bibitem{Martin2008}
\bibinfo{author}{Martin, I.} \& \bibinfo{author}{Batista, C.~D.}
\newblock \bibinfo{title}{{Itinerant Electron-Driven Chiral Magnetic Ordering
  and Spontaneous Quantum Hall Effect in Triangular Lattice Models}}.
\newblock \emph{\bibinfo{journal}{Phys. Rev. Lett.}}
  \textbf{\bibinfo{volume}{101}}, \bibinfo{pages}{156402}
  (\bibinfo{year}{2008}).
\newblock
  \urlprefix\url{https://link.aps.org/doi/10.1103/PhysRevLett.101.156402}.

\bibitem{Akagi2010}
\bibinfo{author}{Akagi, Y.} \& \bibinfo{author}{Motome, Y.}
\newblock \bibinfo{title}{{Spin Chirality Ordering and Anomalous Hall Effect in
  the Ferromagnetic Kondo Lattice Model on a Triangular Lattice}}.
\newblock \emph{\bibinfo{journal}{Journal of the Physical Society of Japan}}
  \textbf{\bibinfo{volume}{79}}, \bibinfo{pages}{083711}
  (\bibinfo{year}{2010}).
\newblock \urlprefix\url{https://doi.org/10.1143/JPSJ.79.083711 d}.

\bibitem{Pasrija2016}
\bibinfo{author}{Pasrija, K.} \& \bibinfo{author}{Kumar, S.}
\newblock \bibinfo{title}{Noncollinear and noncoplanar magnetic order in the
  extended hubbard model on anisotropic triangular lattice}.
\newblock \emph{\bibinfo{journal}{Phys. Rev. B}} \textbf{\bibinfo{volume}{93}},
  \bibinfo{pages}{195110} (\bibinfo{year}{2016}).
\newblock \urlprefix\url{https://link.aps.org/doi/10.1103/PhysRevB.93.195110}.

\bibitem{Pulay1980}
\bibinfo{author}{Pulay, P.}
\newblock \bibinfo{title}{Convergence acceleration of iterative sequences. the
  case of scf iteration}.
\newblock \emph{\bibinfo{journal}{Chemical Physics Letters}}
  \textbf{\bibinfo{volume}{73}}, \bibinfo{pages}{393--398}
  (\bibinfo{year}{1980}).
\newblock
  \urlprefix\url{https://www.sciencedirect.com/science/article/pii/0009261480803964}.

\bibitem{Spalek2007}
\bibinfo{author}{Spalek, J.}
\newblock \bibinfo{title}{{t-J Model Then and Now: a Personal Perspective from
  the Pioneering Times}}.
\newblock \emph{\bibinfo{journal}{Acta Physica Polonica A}}
  \textbf{\bibinfo{volume}{111}}, \bibinfo{pages}{409--424}
  (\bibinfo{year}{2007}).

\bibitem{KeYang2024}
\bibinfo{author}{Yang, K.}, \bibinfo{author}{Chen, Q.}, \bibinfo{author}{Qiao,
  L.} \& \bibinfo{author}{Zhu, Z.}
\newblock \bibinfo{title}{Mean-field study of superconductivity in the
  $t\text{\ensuremath{-}}j$ square lattice model with three-site hopping}.
\newblock \emph{\bibinfo{journal}{Phys. Rev. B}}
  \textbf{\bibinfo{volume}{110}}, \bibinfo{pages}{054514}
  (\bibinfo{year}{2024}).
\newblock \urlprefix\url{https://link.aps.org/doi/10.1103/PhysRevB.110.054514}.

\bibitem{JinChenhao2021}
\bibinfo{author}{Jin, C.} \emph{et~al.}
\newblock \bibinfo{title}{{Stripe phases in WSe2/WS2 moir{\'e} superlattices}}.
\newblock \emph{\bibinfo{journal}{Nature Materials}}
  \textbf{\bibinfo{volume}{20}}, \bibinfo{pages}{940--944}
  (\bibinfo{year}{2021}).
\newblock \urlprefix\url{https://www.nature.com/articles/s41563-021-00959-8}.

\bibitem{XuYang2020}
\bibinfo{author}{Xu, Y.} \emph{et~al.}
\newblock \bibinfo{title}{{Correlated insulating states at fractional fillings
  of moir{\'e} superlattices}}.
\newblock \emph{\bibinfo{journal}{Nature}} \textbf{\bibinfo{volume}{587}},
  \bibinfo{pages}{214--218} (\bibinfo{year}{2020}).
\newblock \urlprefix\url{https://www.nature.com/articles/s41586-020-2868-6}.

\bibitem{Senthil2004Science}
\bibinfo{author}{Senthil, T.}, \bibinfo{author}{Vishwanath, A.},
  \bibinfo{author}{Balents, L.}, \bibinfo{author}{Sachdev, S.} \&
  \bibinfo{author}{Fisher, M. P.~A.}
\newblock \bibinfo{title}{{Deconfined Quantum Critical Points}}.
\newblock \emph{\bibinfo{journal}{Science}} \textbf{\bibinfo{volume}{303}},
  \bibinfo{pages}{1490--1494} (\bibinfo{year}{2004}).
\newblock
  \urlprefix\url{https://www.science.org/doi/abs/10.1126/science.1091806}.

\bibitem{Senthil2004}
\bibinfo{author}{Senthil, T.}, \bibinfo{author}{Balents, L.},
  \bibinfo{author}{Sachdev, S.}, \bibinfo{author}{Vishwanath, A.} \&
  \bibinfo{author}{Fisher, M. P.~A.}
\newblock \bibinfo{title}{{Quantum criticality beyond the
  Landau-Ginzburg-Wilson paradigm}}.
\newblock \emph{\bibinfo{journal}{Phys. Rev. B}} \textbf{\bibinfo{volume}{70}},
  \bibinfo{pages}{144407} (\bibinfo{year}{2004}).
\newblock \urlprefix\url{https://link.aps.org/doi/10.1103/PhysRevB.70.144407}.

\bibitem{Senthil2023}
\bibinfo{author}{Senthil, T.}
\newblock \bibinfo{title}{{Deconfined quantum critical points: a review}}
  (\bibinfo{year}{2023}).
\newblock \eprint{2306.12638}.

\bibitem{LeiWang2020}
\bibinfo{author}{Wang, L.} \emph{et~al.}
\newblock \bibinfo{title}{{Correlated electronic phases in twisted bilayer
  transition metal dichalcogenides}}.
\newblock \emph{\bibinfo{journal}{Nature materials}}
  \textbf{\bibinfo{volume}{19}}, \bibinfo{pages}{861--866}
  (\bibinfo{year}{2020}).

\bibitem{Scholle2023}
\bibinfo{author}{Scholle, R.}, \bibinfo{author}{Bonetti, P.~M.},
  \bibinfo{author}{Vilardi, D.} \& \bibinfo{author}{Metzner, W.}
\newblock \bibinfo{title}{{Comprehensive mean-field analysis of magnetic and
  charge orders in the two-dimensional Hubbard model}}.
\newblock \emph{\bibinfo{journal}{Phys. Rev. B}}
  \textbf{\bibinfo{volume}{108}}, \bibinfo{pages}{035139}
  (\bibinfo{year}{2023}).
\newblock \urlprefix\url{https://link.aps.org/doi/10.1103/PhysRevB.108.035139}.

\bibitem{Riegler2023}
\bibinfo{author}{Riegler, D.} \emph{et~al.}
\newblock \bibinfo{title}{{Interplay of spin and charge order in the
  electron-doped cuprates}}.
\newblock \emph{\bibinfo{journal}{Phys. Rev. B}}
  \textbf{\bibinfo{volume}{108}}, \bibinfo{pages}{195141}
  (\bibinfo{year}{2023}).
\newblock \urlprefix\url{https://link.aps.org/doi/10.1103/PhysRevB.108.195141}.

\bibitem{Scholle2024}
\bibinfo{author}{Scholle, R.}, \bibinfo{author}{Metzner, W.},
  \bibinfo{author}{Vilardi, D.} \& \bibinfo{author}{Bonetti, P.~M.}
\newblock \bibinfo{title}{{Spiral to stripe transition in the two-dimensional
  Hubbard model}}.
\newblock \emph{\bibinfo{journal}{Phys. Rev. B}}
  \textbf{\bibinfo{volume}{109}}, \bibinfo{pages}{235149}
  (\bibinfo{year}{2024}).
\newblock \urlprefix\url{https://link.aps.org/doi/10.1103/PhysRevB.109.235149}.

\end{thebibliography}

\begin{thebibliography}{7}
\expandafter\ifx\csname natexlab\endcsname\relax\def\natexlab#1{#1}\fi
\expandafter\ifx\csname bibnamefont\endcsname\relax
  \def\bibnamefont#1{#1}\fi
\expandafter\ifx\csname bibfnamefont\endcsname\relax
  \def\bibfnamefont#1{#1}\fi
\expandafter\ifx\csname citenamefont\endcsname\relax
  \def\citenamefont#1{#1}\fi
\expandafter\ifx\csname url\endcsname\relax
  \def\url#1{\texttt{#1}}\fi
\expandafter\ifx\csname urlprefix\endcsname\relax\def\urlprefix{URL }\fi
\providecommand{\bibinfo}[2]{#2}
\providecommand{\eprint}[2][]{\url{#2}}

\bibitem[{\citenamefont{Gong et~al.}(2017)\citenamefont{Gong, Li, Li, Ji,
  Stern, Xia, Cao, Bao, Wang, Wang et~al.}}]{ChengGong2017}
\bibinfo{author}{\bibfnamefont{C.}~\bibnamefont{Gong}},
  \bibinfo{author}{\bibfnamefont{L.}~\bibnamefont{Li}},
  \bibinfo{author}{\bibfnamefont{Z.}~\bibnamefont{Li}},
  \bibinfo{author}{\bibfnamefont{H.}~\bibnamefont{Ji}},
  \bibinfo{author}{\bibfnamefont{A.}~\bibnamefont{Stern}},
  \bibinfo{author}{\bibfnamefont{Y.}~\bibnamefont{Xia}},
  \bibinfo{author}{\bibfnamefont{T.}~\bibnamefont{Cao}},
  \bibinfo{author}{\bibfnamefont{W.}~\bibnamefont{Bao}},
  \bibinfo{author}{\bibfnamefont{C.}~\bibnamefont{Wang}},
  \bibinfo{author}{\bibfnamefont{Y.}~\bibnamefont{Wang}}, \bibnamefont{et~al.},
  \bibinfo{journal}{Nature} \textbf{\bibinfo{volume}{546}},
  \bibinfo{pages}{265} (\bibinfo{year}{2017}), ISSN \bibinfo{issn}{1476-4687},
  \urlprefix\url{https://doi.org/10.1038/nature22060}.

\bibitem[{\citenamefont{Bonilla et~al.}(2018)\citenamefont{Bonilla, Kolekar,
  Ma, Diaz, Kalappattil, Das, Eggers, Gutierrez, Phan, and
  Batzill}}]{Bonilla2018}
\bibinfo{author}{\bibfnamefont{M.}~\bibnamefont{Bonilla}},
  \bibinfo{author}{\bibfnamefont{S.}~\bibnamefont{Kolekar}},
  \bibinfo{author}{\bibfnamefont{Y.}~\bibnamefont{Ma}},
  \bibinfo{author}{\bibfnamefont{H.~C.} \bibnamefont{Diaz}},
  \bibinfo{author}{\bibfnamefont{V.}~\bibnamefont{Kalappattil}},
  \bibinfo{author}{\bibfnamefont{R.}~\bibnamefont{Das}},
  \bibinfo{author}{\bibfnamefont{T.}~\bibnamefont{Eggers}},
  \bibinfo{author}{\bibfnamefont{H.~R.} \bibnamefont{Gutierrez}},
  \bibinfo{author}{\bibfnamefont{M.-H.} \bibnamefont{Phan}}, \bibnamefont{and}
  \bibinfo{author}{\bibfnamefont{M.}~\bibnamefont{Batzill}},
  \bibinfo{journal}{Nature Nanotechnology} \textbf{\bibinfo{volume}{13}},
  \bibinfo{pages}{289} (\bibinfo{year}{2018}), ISSN \bibinfo{issn}{1748-3387},
  \urlprefix\url{https://www.nature.com/articles/s41565-018-0063-9}.

\bibitem[{\citenamefont{Deng et~al.}(2018)\citenamefont{Deng, Yu, Song, Zhang,
  Wang, Sun, Yi, Wu, Wu, Zhu et~al.}}]{YujunDeng2018}
\bibinfo{author}{\bibfnamefont{Y.}~\bibnamefont{Deng}},
  \bibinfo{author}{\bibfnamefont{Y.}~\bibnamefont{Yu}},
  \bibinfo{author}{\bibfnamefont{Y.}~\bibnamefont{Song}},
  \bibinfo{author}{\bibfnamefont{J.}~\bibnamefont{Zhang}},
  \bibinfo{author}{\bibfnamefont{N.~Z.} \bibnamefont{Wang}},
  \bibinfo{author}{\bibfnamefont{Z.}~\bibnamefont{Sun}},
  \bibinfo{author}{\bibfnamefont{Y.}~\bibnamefont{Yi}},
  \bibinfo{author}{\bibfnamefont{Y.~Z.} \bibnamefont{Wu}},
  \bibinfo{author}{\bibfnamefont{S.}~\bibnamefont{Wu}},
  \bibinfo{author}{\bibfnamefont{J.}~\bibnamefont{Zhu}}, \bibnamefont{et~al.},
  \bibinfo{journal}{Nature} \textbf{\bibinfo{volume}{563}}, \bibinfo{pages}{94}
  (\bibinfo{year}{2018}), ISSN \bibinfo{issn}{0028-0836},
  \urlprefix\url{https://www.nature.com/articles/s41586-018-0626-9}.

\bibitem[{\citenamefont{Gibertini et~al.}(2019)\citenamefont{Gibertini,
  Koperski, Morpurgo, and Novoselov}}]{Gibertini2019}
\bibinfo{author}{\bibfnamefont{M.}~\bibnamefont{Gibertini}},
  \bibinfo{author}{\bibfnamefont{M.}~\bibnamefont{Koperski}},
  \bibinfo{author}{\bibfnamefont{A.~F.} \bibnamefont{Morpurgo}},
  \bibnamefont{and} \bibinfo{author}{\bibfnamefont{K.~S.}
  \bibnamefont{Novoselov}}, \bibinfo{journal}{Nature Nanotechnology}
  \textbf{\bibinfo{volume}{14}}, \bibinfo{pages}{408} (\bibinfo{year}{2019}),
  ISSN \bibinfo{issn}{1748-3387},
  \urlprefix\url{https://www.nature.com/articles/s41565-019-0438-6}.

\bibitem[{\citenamefont{Gong and Zhang}(2019)}]{ChengGong2019}
\bibinfo{author}{\bibfnamefont{C.}~\bibnamefont{Gong}} \bibnamefont{and}
  \bibinfo{author}{\bibfnamefont{X.}~\bibnamefont{Zhang}},
  \bibinfo{journal}{Science} \textbf{\bibinfo{volume}{363}},
  \bibinfo{pages}{eaav4450} (\bibinfo{year}{2019}),
  \urlprefix\url{https://www.science.org/doi/abs/10.1126/science.aav4450}.

\bibitem[{\citenamefont{Bedoya-Pinto et~al.}(2021)\citenamefont{Bedoya-Pinto,
  Ji, Pandeya, Gargiani, Valvidares, Sessi, Taylor, Radu, Chang, and
  Parkin}}]{BedoyaAmilcar2021}
\bibinfo{author}{\bibfnamefont{A.}~\bibnamefont{Bedoya-Pinto}},
  \bibinfo{author}{\bibfnamefont{J.-R.} \bibnamefont{Ji}},
  \bibinfo{author}{\bibfnamefont{A.~K.} \bibnamefont{Pandeya}},
  \bibinfo{author}{\bibfnamefont{P.}~\bibnamefont{Gargiani}},
  \bibinfo{author}{\bibfnamefont{M.}~\bibnamefont{Valvidares}},
  \bibinfo{author}{\bibfnamefont{P.}~\bibnamefont{Sessi}},
  \bibinfo{author}{\bibfnamefont{J.~M.} \bibnamefont{Taylor}},
  \bibinfo{author}{\bibfnamefont{F.}~\bibnamefont{Radu}},
  \bibinfo{author}{\bibfnamefont{K.}~\bibnamefont{Chang}}, \bibnamefont{and}
  \bibinfo{author}{\bibfnamefont{S.~S.~P.} \bibnamefont{Parkin}},
  \bibinfo{journal}{Science} \textbf{\bibinfo{volume}{374}},
  \bibinfo{pages}{616} (\bibinfo{year}{2021}),
  \urlprefix\url{https://www.science.org/doi/abs/10.1126/science.abd5146}.

\bibitem[{\citenamefont{Wang et~al.}(2022)\citenamefont{Wang, Li, Wen, Cheng,
  Yin, Liu, Li, and He}}]{HaoWang2022}
\bibinfo{author}{\bibfnamefont{H.}~\bibnamefont{Wang}},
  \bibinfo{author}{\bibfnamefont{X.}~\bibnamefont{Li}},
  \bibinfo{author}{\bibfnamefont{Y.}~\bibnamefont{Wen}},
  \bibinfo{author}{\bibfnamefont{R.}~\bibnamefont{Cheng}},
  \bibinfo{author}{\bibfnamefont{L.}~\bibnamefont{Yin}},
  \bibinfo{author}{\bibfnamefont{C.}~\bibnamefont{Liu}},
  \bibinfo{author}{\bibfnamefont{Z.}~\bibnamefont{Li}}, \bibnamefont{and}
  \bibinfo{author}{\bibfnamefont{J.}~\bibnamefont{He}},
  \bibinfo{journal}{Applied Physics Letters} \textbf{\bibinfo{volume}{121}},
  \bibinfo{pages}{220501} (\bibinfo{year}{2022}), ISSN
  \bibinfo{issn}{0003-6951}, \urlprefix\url{https://doi.org/10.1063/5.0130037}.

\end{thebibliography}
\end{document}